\documentclass[journal=jpccck,manuscript=article]{achemso}
\setkeys{acs}{articletitle = true}
\usepackage[version=3]{mhchem} 
\usepackage{color}
\usepackage{amsfonts}
\usepackage{amssymb}
\usepackage{amsmath}
\usepackage{graphicx}
\usepackage{float}
\author{Shaul Mukamel}
\email{smukamel@uci.edu}
\phone{+1 949 824 7600}
\affiliation{Department of Chemistry, University of California Irvine, Irvine, CA 92697, USA}
\author{Michael Galperin}
\email{migalperin@ucsd.edu}
\phone{+1 858 246 0511}
\affiliation{Department of Chemistry and Biochemistry, University of California San Diego, La Jolla, CA 92093, USA}
\title{Flux-conserving diagrammatic formulation of optical spectroscopy of open quantum systems}
\begin{document}
\begin{abstract}
We present a theoretical approach to optical spectroscopy of open nonequilibrium systems,
which generalizes traditional nonlinear optical spectroscopy tools by imposing charge and
energy conservation at all levels of approximation.
Both molecular and radiation field degrees of freedom are treated quantum 
mechanically. 
The formulation is based on the nonequilibrium Green's function (NEGF) approach
and a double sided Feynman  diagrammatic representation of the photon flux is developed.
Numerical simulations are presented for a model system.
Our study  bridges the theoretical approaches of quantum transport and 
optical spectroscopy and establishes a firm basis for applying traditional tools
of nonlinear optical spectroscopy in molecular optoelectronics.
\end{abstract}

\maketitle


\section{Introduction}\label{intro}
Optical spectroscopy is a standard tool for probing and controlling electronic and vibrational structure 
and dynamics in molecular systems. For example, attosecond pulses make real-time 
observation of atomic scale electron dynamics possible~\cite{krausz_attosecond_2009},
localized surface plasmons allow to go beyond diffraction limit achieving single-molecule
sensitivity~\cite{kawata_plasmonics_2009}, surface and tip enhanced Raman spectroscopy
yields information on single molecule vibrational structure and excitations~\cite{le_ru_single-molecule_2012,verma_tip-enhanced_2017},
tetrahertz electromagnetic radiation  provides access to rotational degrees of freedom of molecules~\cite{kampfrath_resonant_2013}
and X-ray spectroscopy gives access to electronic transitions and nuclear dynamics~\cite{young_roadmap_2018,bennett_monitoring_2018}.
Recently, quantum effects of radiation have attracted attention as well~\cite{schlawin_entangled_2018,dorfman_monitoring_2019,asban_quantum_2019}.

Advances in nanoscale fabrication techniques allow optical 
measurements in current-carrying single-molecule junctions.
In particular, bias-induced luminescence was used to observe vibrationally resolved features with sub-molecular precision~\cite{qiu_vibrationally_2003,dong_vibrationally_2004,HoPRB08,chen_viewing_2010},
visualize inter-molecular dipole-dipole coupling~\cite{zhang_visualizing_2016},
investigate energy transfer in molecular dimers~\cite{imada_real-space_2016},
study selective triplet exciton formation in single molecule~\cite{kimura_selective_2019},
and to access information on electronic quantum shot noise in the junction~\cite{BerndtPRL12}.
Raman spectroscopy was utilized to resolve bias-dependent vibrational fingerprint of a molecule in
a junction~\cite{natelson_nanogap_2013}, 
to observe time-dependent correlations between conductance and optical signal~\cite{NatelsonNL08},
and to estimate extent of bias-induced vibrational and electronic heating in junctions~\cite{CheshnovskySelzerNatNano08,NatelsonNatNano11}.
Optical read-out of the junction response to nanosecond voltage pulses was utilized
to enable access to transient processes~\cite{LothAPL13}.
Performing and interpreting optical experiments in open non-equilibrium molecular systems
requires the combination of two research areas - optical spectroscopy and molecular electronics -
indicating the emergence of a new research direction coined molecular optoelectronics~\cite{MGANPCCP12}.

The theory of nonlinear optical spectroscopy of molecules is well established~\cite{OkumuraJCP97,OkumuraTanimuraJCP97,FlemingJPCA01,OvchinnikovApkarianVothJCP01,OkumuraJPCA03,MukamelChemRev04,MukamelPRA05,MukamelPRB08,MukamelPRB09}.
A unifying framework for the interpretation of optical measurements in molecules
was published in the book ``Principles of Nonlinear Optical Spectroscopy''~\cite{Mukamel_1995},
the very classification of ultrafast optical processes is based on
{\em double-sided Feynman diagrams} first introduced in the book.
These represent a bare perturbation theory expansion of the molecular density matrix
in light-matter interaction. Caution should be exercised with the approach in open systems. 
When the radiaton field is treated classically the bare 
perturbation expansion holds for closed and open systems alike, and the double-sided
Feynman diagrams are constructed in the usual way. The only restriction is the necessity
to avoid quantum regression statement~\cite{breuer_theory_2003}, when evaluating multi-time
correlation functions of electronic operators~\cite{gao_simulation_2016}.
However, the treatment of quantum radiation fields is more involved~\cite{galperin_photonics_2017}.
Mutual influence of two quantum subsystems (e.g., radiation field and electronic degrees of freedom)
leads to restrictions on building perturbative expansions in their interaction~\cite{paper_663}.
Bare perturbation theory for quantum light in open systems does not conserve
charge and energy~\cite{BaymKadanoffPR61,BaymPR62,kadanoff_quantum_1962,gao_optical_2016}
and may even lead to qualitative failures due to lack of account for back action 
from one system on the other in the bare expansion~\cite{nitzan_kinetic_2018}.

Here, we develop a Green's function approach whereby charge and energy conservation are built in.
That is, total charge and total energy in the whole system do not change during the evolution.
A double sided Feynman diagrammatic expansion of  the Green's functions that can describe  the response 
of open systems to quantum fields in terms of pathways is developed.

The structure of the paper is as follows. After introducing a model of open system
subjected to quantum radiation,
we consider consider diagrammatic expansion in light-matter interaction in 
and discuss possible generalization of
double-sided Feynman diagrams. 
Theoretical discussion is followed by illustrative numerical simulations.
We conclude with short summary and directions for future research. 

\section{Theoretical Methods}
\subsection{Model}\label{model}
We consider a junction consisting of molecule $M$ coupled to two metallic contacts 
$L$ and $R$ (each at its own equilibrium) and to external quantum radiation field modes.
The system Hamiltonian is
\begin{align}
\label{H}
\hat H &= \hat H_0 + \hat V
\\
\hat H_0 &= \hat H_M + \hat H_L + \hat H_R + \hat H_{rad}
\\
\label{V}
\hat V &= \hat V_{ML} + \hat V_{MR} + \hat V_{M,rad}
\end{align}
where $\hat H_0$ represents uncoupled molecule ($\hat H_M$), contacts ($\hat H_{L}$ and $\hat H_{R}$),
and radiation field ($\hat H_{rad}$), while $\hat V$ gives the interaction between the sub-systems.
The molecular Hamiltonian $\hat H_M$ is assumed to be quadratic in the fermi operators (neglecting electron correlations),
the contacts are modeled as continua of free charge carriers, the radiation field is expanded in
a set of modes.
\begin{align}
\label{HM}
\hat H_M &= \sum_{m_1,m_2} H^{M}_{m_2m_2} \hat d_{m_1}^\dagger\hat d_{m_2}
\\
\label{HK}
 \hat H_K &= \sum_{k\in K}\varepsilon_k\hat c_k^\dagger\hat c_k
 \\
 \label{Hrad}
 \hat H_{rad} &= \sum_{\alpha} \omega_\alpha\hat a_\alpha^\dagger\hat a_\alpha
 \\
 \label{VMK}
 \hat V_{MK} &= \sum_{m\in M}\sum_{k\in K}\bigg( V_{mk}\hat d^\dagger_m\hat c_k + H.c.\bigg)
 \\
 \label{VMrad}
 \hat V_{M,rad} &= \sum_{m_1,m_2\in M}\sum_{\alpha}
\left(U_{\alpha,m_1m_2}\,\hat a_\alpha^\dagger\hat D_{m_1m_2} + H.c.\right)
\end{align}
Here  $\hat d_m^\dagger$ ($\hat d_m$) and $\hat c_k^\dagger$ ($\hat c_k$)
 create (annihilate) electron in the molecular orbital $m$ or orbital $k$ of the contacts,
 respectively. $\hat D_{m_1m_2}\equiv \hat d_{m_1}^\dagger\hat d_{m_2}$ 
 is the molecular de-excitation operator. $\hat a_\alpha^\dagger$ ($\hat a_\alpha$)
 creates (annihilates) a photon in mode $\alpha$ of the radiation field.
 
 We shall develop systematic approximations for electron and photon fluxes
 defined as the rate of change of population in contacts and radiation field respectively
 that conserve the fluxes, i.e. the charge and energy of the entire system does not change during evolution.
 \begin{align}
 \label{IKdef}
 I_K(t) &\equiv -\frac{d}{dt}\sum_{k\in K} \langle \hat c_k^\dagger(t)\hat c_k(t)\rangle
 \qquad (K=L,R)
 \\
 \label{Iptdef}
 I_{pt}(t) &\equiv +\frac{d}{dt}\sum_{\alpha} \langle \hat a_\alpha^\dagger(t)\hat a_{\alpha} \rangle
 \end{align}
 and corresponding energy fluxes defined as rate of change of energy
 \begin{align}
 \label{JKdef}
 J_K(t) &\equiv -\frac{d}{dt}\sum_{k\in K} \varepsilon_k \langle \hat c_k^\dagger(t)\hat c_k(t)\rangle
 \qquad (K=L,R)
 \\
 \label{Jptdef}
 J_{pt}(t) &\equiv +\frac{d}{dt}\sum_{\alpha} \omega_\alpha \langle \hat a_\alpha^\dagger(t)\hat a_{\alpha} \rangle
 \end{align}
We adopt the conventional notation in quantum transport whereby positive electron flux is the flux from bath (contact)
 into system (molecule), while in optical spectroscopy positive photon flux goes from system (molecule)
  into bath (radiation field modes).
 

\subsection{Expanding the fluxes in the light-matter interaction}\label{weak}
The perturbative expansion is developed around the zero-order Hamiltonian $\hat H_0$.
Standard nonequilibrium Green's function theory (NEGF) aims at calculating the
electron and photon Green's functions defined on the Keldysh contour in the Heisenberg picture
\begin{align}
\label{Gdef}
 G_{m_1m_2}(\tau_1,\tau_2) &\equiv -i\langle T_c\,\hat d_{m_1}(\tau_1)\,\hat d_{m_2}^\dagger(\tau_2)\rangle
 \\
 \label{Fdef}
 F_{\alpha_1\alpha_2}(\tau_1,\tau_2) &\equiv -i\langle T_c\,\hat a_{\alpha_1}(\tau_1)\,\hat a_{\alpha_2}^\dagger(\tau_2)\rangle
\end{align}
These satisfy set of {\em exact} coupled Dyson equations~\cite{HaugJauho_2008,StefanucciVanLeeuwen_2013}
\begin{align}
\label{DysonG}
 &\sum_m\int_c d\tau\bigg[
 \delta(\tau_1,\tau)\bigg(i \delta_{m_1,m}\frac{\partial}{\partial \tau} - H^{M}_{m_1m}\bigg)
 -\sum_{K=L,R}\Sigma^{K}_{m_1m}(\tau_1,\tau)\bigg] G_{mm_2}(\tau,\tau_2) 
 \\ &\qquad\qquad\qquad
 = 
 \delta_{m_1,m_2}\delta(\tau_1,\tau_2) + \sum_m\int_cd\tau\, \Sigma^{pt}_{m_1m}(\tau_1,\tau)\,
 G_{mm_2}(\tau,\tau_2)
 \nonumber \\
 \label{DysonF}
 &\bigg( i\frac{\partial}{\partial\tau_1} -\omega_{\alpha_1} \bigg) F_{\alpha_1\alpha_2}(\tau_1,\tau_2)
 = \delta_{\alpha_1,\alpha_2}\delta(\tau_1,\tau_2)
 + \sum_\alpha\int_cd\tau\, \Pi^{el}_{\alpha_1\alpha}(\tau_1,\tau)\, F_{\alpha\alpha_2}(\tau,\tau_2)
\end{align}
Here $\Sigma^{K}$ ($K=L,R$), $\Sigma^{pt}$ and $\Pi^{el}$ are self-energies of
electrons due to coupling to contact $K$, electrons due to coupling to radiation field modes, 
and photons due to coupling to the electronic subsystem.

The bilinear molecule-contacts coupling, eq~\ref{VMK}, results in an exact expression for the self-energy $\Sigma^K$
\begin{equation}
\label{SigmaK}
 \Sigma^K_{m_1m_2}(\tau_1,\tau_2) = \sum_{k\in K} V_{m_1k}\, g_k(\tau_1,\tau_2)\, V_{km_2},
\end{equation}
where $g_k(\tau_1,\tau_2) \equiv -i\langle T_c\,\hat c_k(\tau_1)\,\hat c_k^\dagger(\tau_2)\rangle$
is Green's function for free electrons in state $k$ of contact $K$. 
Its projections are $g_k^r(t_1,t_2)=-i\theta(t_1-t_2)\,e^{-i\varepsilon_k(t_1-t_2)}$,
$g_k^<(t_1,t_2)=i\, n_k\,e^{-i\varepsilon_k(t_1-t_2)}$, $g_k^>(t_1,t_2)=-i[1-n_k]\,e^{-i\varepsilon_k(t_1-t_2)}$.
The self-energy projections in the frequency domain are ($K=L,R$)
\begin{align}
\Sigma^{K\, r}_{m_1m_2}(E) &= \Lambda^K_{m_1m_2}(E) - \frac{i}{2}\Gamma^K_{m_1m_2}(E)
\\
\Sigma^{K\, <}_{m_1m_2}(E) &= i\,\Gamma^K_{m_1m_2}(E)\, f_K(E)
\\
\Sigma^{K\, >}_{m_1m_2}(E) &= -i\,\Gamma^K_{m_1m_2}(E)\,[1-f_K(E)]
\end{align}
Here $r$, $<$ and $>$ superscripts indicate retarded, lesser and greater projections, $f_K(E)$ is the Fermi-Dirac
thermal distribution in the contacts.
\begin{equation}
\Gamma^K_{m_1m_2}(E) \equiv 2\pi\sum_{k\in K} V_{m_1k}\, V_{km_2}\,\delta(E-\varepsilon_k)
\end{equation}
is a dissipation matrix due to coupling to contact $K$, and $\Lambda^K$ is the Lamb shift related to to $\Gamma^K$ via
the Kramers-Kronig relations. 

\begin{figure}[htbp]
\centering\includegraphics[width=0.8\linewidth]{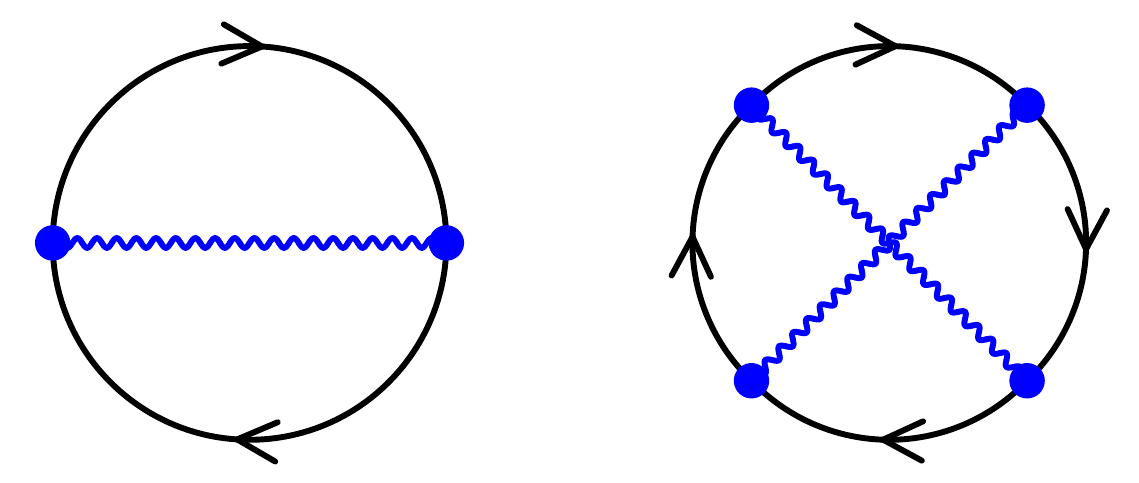}
\caption{\label{fig1}
Diagrammatic perturbation theory within NEGF.
Shown are the Luttinger-Ward generating functionals for second (left) and fourth (right)
order expansion in light-matter interaction.
Directed solid line (black) represents the electron Green function $G$, eq~\ref{Gdef}.
Wavy line (blue) is the photon Green function $F$, eq~\ref{Fdef}; 
both directions are implied here. Solid circles indicate vertices.
Summation over all degrees of freedom and integration 
over contour variables is  assumed at the vertices.
}
\end{figure}

$\Sigma^{pt}$ and $\Pi^{el}$ must be calculated approximately.
Within the NEGF self-energies are defined as functional derivatives of  the Luttinger-Ward functional 
$\Phi$~\cite{LuttingerWardPR60,Haussmann_1999,StefanucciVanLeeuwen_2013}
(see, e.g., eq~3.12 in Ref.~\citenum{Haussmann_1999})
\begin{align}
\label{SigmaptPhi}
 \Sigma^{pt}_{m_1m_2}(\tau_1,\tau_2) &= +\frac{\delta\Phi}{\delta G_{m_2m_1}(\tau_2,\tau_1)}
 \\
 \label{PielPhi}
 \Pi^{el}_{\alpha_1\alpha_2}(\tau_1,\tau_2) &= -\frac{\delta\Phi}{\delta F_{\alpha_2\alpha_1}(\tau_2,\tau_1)}
\end{align}
Diagrams for the Luttinger-Ward functional to fourth order in light-matter interaction $\hat V_{M,rad}$, eq~\ref{VMrad}, 
are shown in Figure~\ref{fig1}.
\begin{align}
\Phi &= i\sum_{\{\alpha\}}\sum_{\{m\}}\int_cd\tau_1\int_cd\tau_2\, 
U_{m_1m_2,\alpha_1}\,F_{\alpha_1\alpha_2}(\tau_1,\tau_2)\, U_{\alpha_2,m_3m_4}\, 
G_{m_1m_3}(\tau_1,\tau_2)\, G_{m_4m_2}(\tau_2,\tau_1)
\nonumber\\
& - \sum_{\{\alpha\}}\sum_{\{m\}}\int_cd\tau_1\int_cd\tau_2\int_cd\tau_3\int_cd\tau_4\,
U_{m_1m_2,\alpha_1}\,F_{\alpha_1\alpha_3}(\tau_1,\tau_3)\, U_{\alpha_3,m_3m_4}\, 
 \\ &\times
U_{m_5m_6,\alpha_2}\,F_{\alpha_2\alpha_4}(\tau_2,\tau_4)\, U_{\alpha_4,m_7m_8}\, 
G_{m_1m_6}(\tau_1,\tau_2)\, G_{m_5m_3}(\tau_2,\tau_3)\, G_{m_4m_7}(\tau_3,\tau_4)\, G_{m_8m_2}(\tau_4,\tau_1)
\nonumber
\end{align}
Self-energies constructed in this way are known to preserve all
conservation laws in each order~\cite{BaymKadanoffPR61,BaymPR62,kadanoff_quantum_1962}.
Explicit expressions for the self-energies to fourth order in $\hat V_{M,rad}$ are given in the Supporting Information.

We note that the diagrammatic expansion is performed in the entire $V$, eq~\ref{V},
which includes both molecule-contacts and molecule-radiation field couplings.
However, since the coupling to the contacts, eq~\ref{VMK}, is quadratic, it is
{\em exactly} resummed into the self-energy $\Sigma^K$, eq~\ref{SigmaK}, while 
the molecule-radiation interaction can be accounted for through a perturbative expansion in the light-matter interaction.

Computing the Green's functions and self-energies is a bit different for time-dependent and steady-state applications. 
In the former case one has to solve time-dependent problem, which consists of setting initial conditions for 
the Green's functions. Because of causality self-energies required for a particular time step only depend on Green's
functions at earlier times. So that starting from an initial condition one is able to propagate equations of motion step-by-step. 
Details of time propagation were discussed in, e.g., Ref.~\citenum{stan_time_2009}. 
Note that the initial condition may include either decoupled system and baths (contacts and radiation field) 
with sudden or  adiabatic switching of the coupling, or steady-state junction (coupling to contacts switched at infinite past) 
subjected at time $t_0$ to laser pulse.
Also propagation on two-dimensional time grid is extremely heavy numerically,
so that approximate schemes reducing to a single time propagation were developed~\cite{latini_charge_2014}.

For steady-state, coupling to both contacts and radiation field are assumed to happen
in the infinite past, and particular form of the switching (sudden or adiabatic) is not important because by the time 
steady-state was established, transients die out. In this case we Fourier transform the Dyson equations, 
Green's functions and self-energies to energy space.
Dyson equations, eqs~\ref{DysonG}-\ref{DysonF}, with self-energies,  eqs~S1-S2,
have to be solved self-consistently until convergence starting from Green's function for, e.g., decoupled electronic
and photon systems.
In summary, such procedure consists of the following steps:
\begin{enumerate}
\item Obtain Green's functions for decoupled electrons and photons (e.g., solve problem for molecular junction in the absence of the field to get electron Green's function and assume free photon field - e.g., CW laser, for photon Green's function).
\item\label{step2}
Use the Green's functions to evaluate the self-energies, eqs~S1-S2.
\item Use the self-energies to calculate Green's functions by numerically solving the Dyson equations, eqs~\ref{DysonG}-\ref{DysonF}.
\item Check convergence by, e.g., calculating populations of electronic levels and photon modes.
If difference on two steps of the procedure is less than predefined tolerance, stop the calculation; otherwise
return to step \ref{step2}.
\end{enumerate}

Once the self-energies and Green's functions are known, one can calculate the fluxes, eqs~\ref{IKdef}-\ref{Jptdef}.
Within NEGF {\em exact} expressions for the fluxes are obtained by following the celebrated Jauho-Wingreen-Meir derivation~\cite{jauho_time-dependent_1994,MGNitzanRatner_heat_PRB07,HaugJauho_2008}
\begin{align}
\label{IK_NEGF}
I_K(t) &= 2\,\mbox{Re}\int_{-\infty}^t dt'\,\mbox{Tr}
\bigg[\Sigma^{K\, <}(t,t')\, G^{>}(t',t) - \Sigma^{K\, >}(t,t')\, G^{<}(t',t)\bigg]
\\
\label{Ipt_NEGF}
I_{pt}(t) &= 2\,\mbox{Re}\int_{-\infty}^t dt'\,\mbox{Tr}
\bigg[F^{<}(t,t')\,\Pi^{el\, >}(t',t)- F^{>}(t,t')\, \Pi^{el\, <}(t',t)\bigg]
\\
\label{JK_NEGF}
J_K(t) &= 2\,\mbox{Im}\int_{-\infty}^t dt'\,\mbox{Tr}
\bigg[\frac{\partial\Sigma^{K\, >}(t,t')}{\partial t}\, G^{<}(t',t) 
       - \frac{\partial\Sigma^{K\, <}(t,t')}{\partial t}\, G^{>}(t',t)\bigg]
\\
\label{Jpt_NEGF}
J_{pt}(t) &= 2\,\mbox{Im}\int_{-\infty}^t dt'\,\mbox{Tr}
\bigg[\frac{\partial F^{>}(t,t')}{\partial t} \Pi^{el\, <}(t',t)
- \frac{\partial \bigg[F^{<}(t,t')}{\partial t} \Pi^{el\, >}(t',t) \bigg]
\\
& +2\,\mbox{Re}\int_{-\infty}^t dt'\int_{-\infty}^{t'} dt''\, \mbox{Tr}
\bigg[\Pi^{el\, <}(t,t'')\, F^{>}(t'',t')\,\Pi^{el\, >}(t',t) 
\nonumber \\ & \qquad\qquad\qquad\qquad\qquad\ \,
+ \Pi^{el\, >}(t,t'')\, F^{<}(t'',t')\,\Pi^{el\, <}(t',t) \bigg]
\nonumber
\end{align}
Here the trace is over molecular orbitals in eqs~\ref{IK_NEGF} and \ref{JK_NEGF}
and over radiation field modes in eqs~\ref{Ipt_NEGF} and \ref{Jpt_NEGF}.
Note that the fluxes are coupled, because the self-energies
entering their definitions are derived form the same Luttinger-Ward functional
(see Figure~\ref{fig1}). Thus, they should be treated on equal footing.
This interdependence of fluxes results in charge and energy conservation
(see below for a simple illustration). 
Note that in the usual NEGF approach the molecule-contacts coupling is switched on
at the infinite past - thus minus infinity as lower limit in integrals in eqs~\ref{IK_NEGF}-\ref{Jpt_NEGF}.
However, other switchings are possible.

\section{Results and Discussion}
\subsection{Double-sided Feynman diagrams for the Green's functions}\label{diag}

\begin{figure}[htbp]
\centering\includegraphics[width=0.8\linewidth]{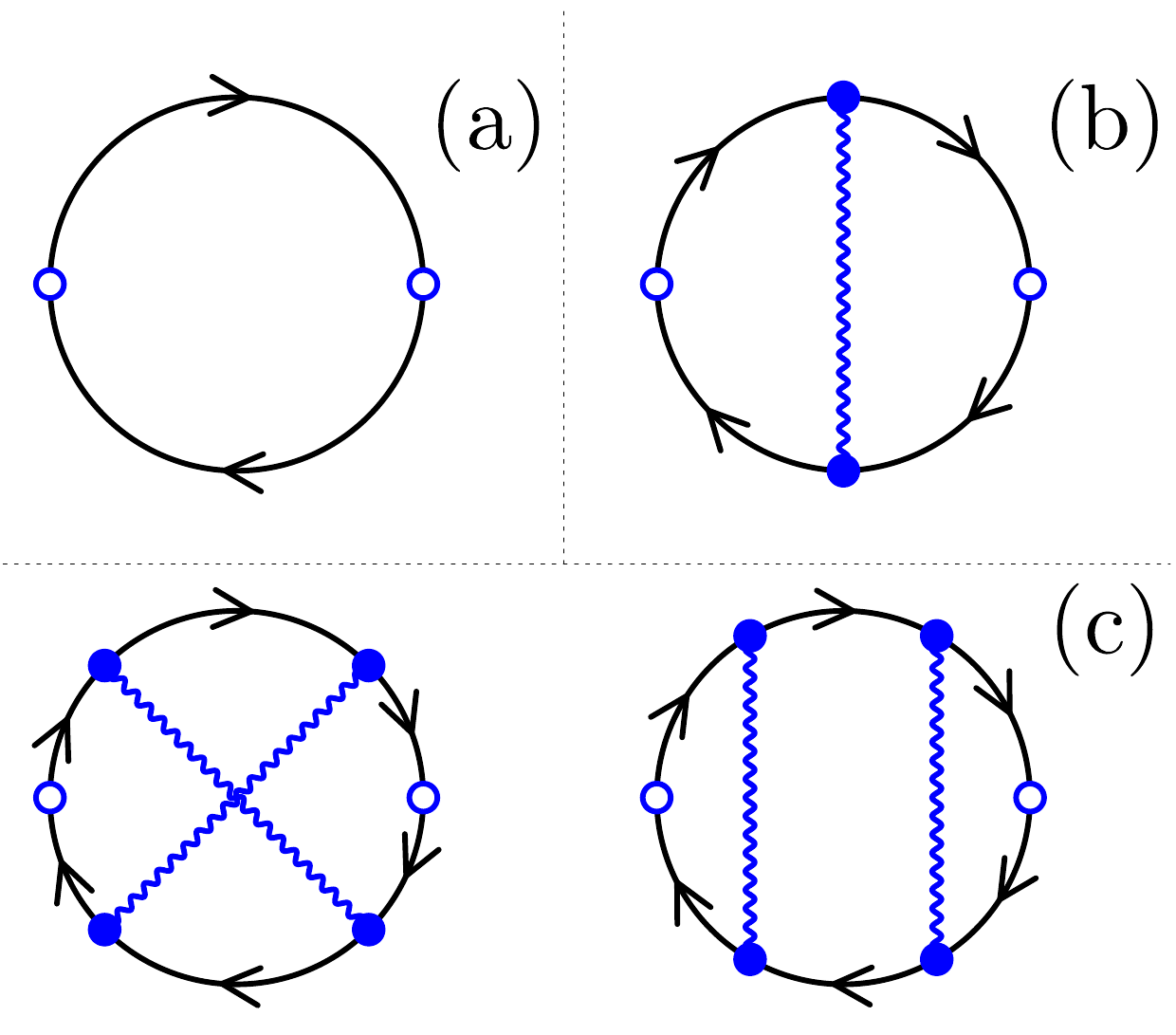}
\caption{\label{fig2}
Diagrams contributing to photon self-energy due to coupling to electrons $\Pi^{el}$.
Shown are contributions of (a) second, (b) fourth, and (c) sixth orders.
Directed solid line (black) represents the electron Green function $G$, eq~\ref{Gdef}.
Wavy line (blue) is the photon Green function $F$, eq~\ref{Fdef}; 
both directions are implied here. Open and solid circles indicate outer and inner vertices.
Summation over all degrees of freedom and integration over contour variables is assumed
for inner vertices.
}
\end{figure}

Below we present a double-sided Feynman diagram
expansion of the fluxes, based on the diagrammatic expansion of the self-energies.
It is important to stress the difference in language between Green's function (Hilbert space) and
density matrix (Liouville space) formulations. 
Original double-sided Feynman diagrams act in Liouville space.
Corresponding construction in the Hilbert space within Green's function technique
is called projection, while term diagram is reserved for representation of 
irreducible contributions within perturbative expansion.
Figure~\ref{fig2} shows second (a), fourth (b),
and sixth (c) order Feynman diagrams contributing to photon self-energy due to coupling to electrons, $\Pi^{el}$.
Each diagram can be projected on the Keldysh contour resulting in a set of 
contributions, which in Liouville space language are denoted double-sided Feynman diagrams. 
An important point is that while in second and fourth order, where only one
diagram contributes to the self-energy, difference in the languages is of secondary importance,
one has to be careful with sixth order contribution, where
two different diagrams (see Figure~\ref{fig2}c) representing different physical processes 
will have same set of time projections.   
Another difference to keep in mind is time ordering in the two approaches:
while Green's function projections only account for ordering along the Keldysh contour,
Liouville space formulation requires also ordering in physical time. Thus, one Hilbert space
projection represents several Liouville space diagrams
(see, e.g., Ref.~\citenum{bergmann_electron_2019} for more details).

\begin{figure}[htbp]
\centering\includegraphics[width=0.8\linewidth]{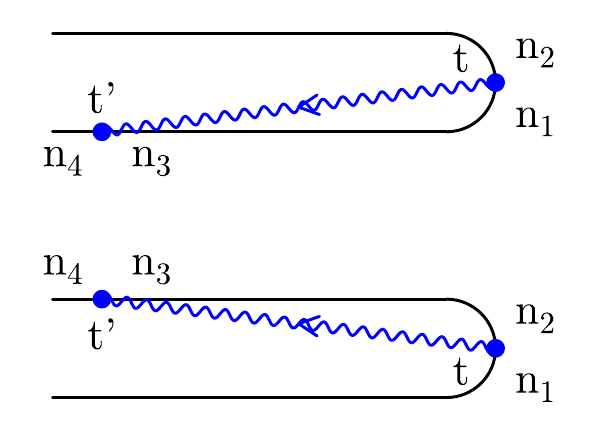}
\caption{\label{fig3}
Double-sided Feynman diagrams for second order optical processes in the photon flux, eq~\ref{Ipt2}.
Wavy line (blue) is the photon Green function $F$, eq~\ref{Fdef}.
Top (bottom) diagram corresponds to first (second) term in the right side of eq~\ref{Ipt2}.
Indices $n_i$ indicate molecular orbitals.
}
\end{figure}

\begin{figure}[htbp]
\centering\includegraphics[width=0.8\linewidth]{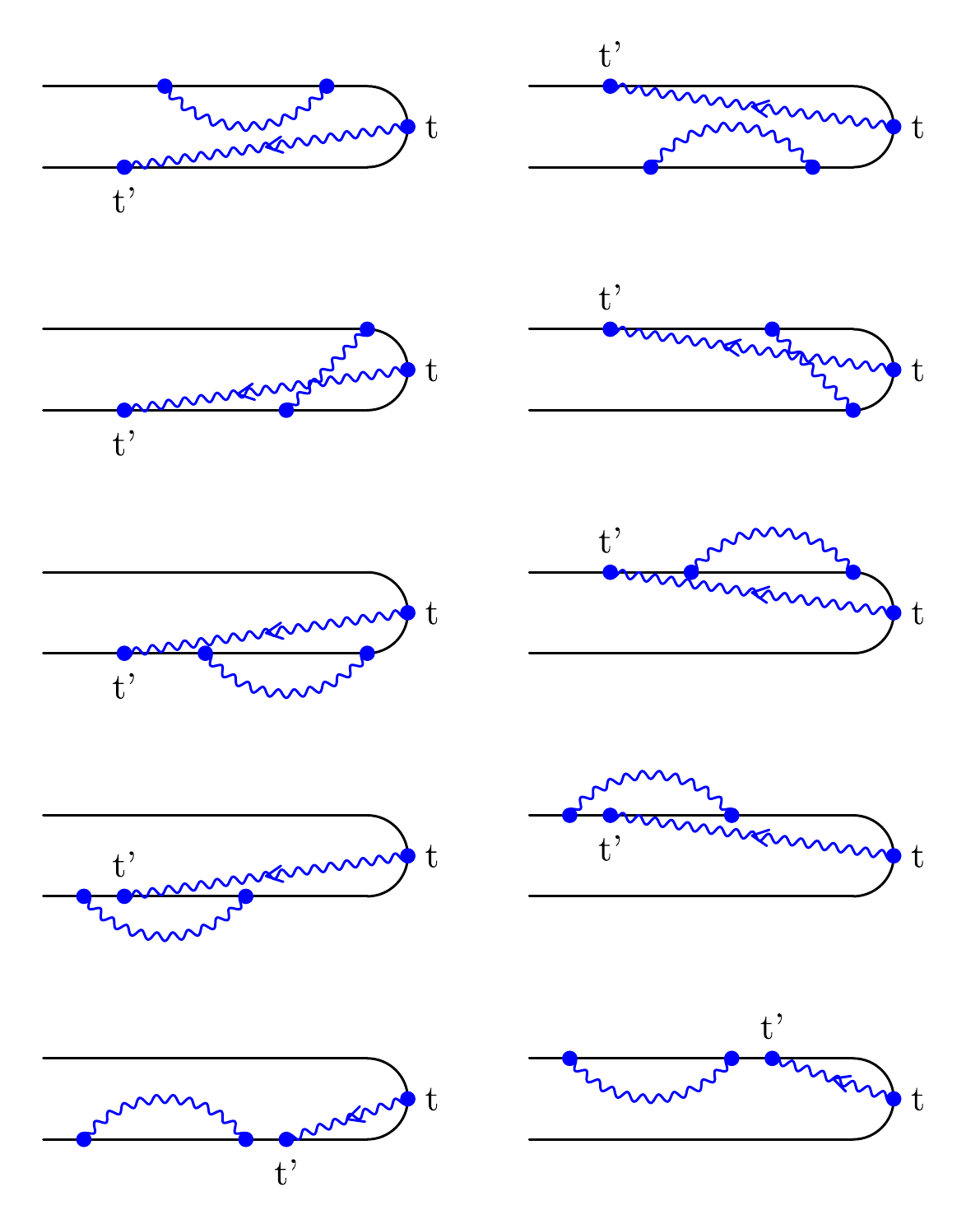}
\caption{\label{fig4}
Double-sided Feynman diagrams for fourth order optical processes in the photon flux.
Wavy line (blue) is the photon Green's function $F$, eq~\ref{Fdef}.
Left (right) column corresponds to first (second) term in the right side of eq~\ref{Ipt_NEGF}.
Wavy line without arrow stands for $F(\tau_3,\tau_4)$ of eq~S2.
Both arrow directions are possible in this line.
}
\end{figure}

We are now ready to introduce double-sided Feynman diagrams for the photon flux, eq~\ref{Ipt_NEGF}.
Indeed, double-sided Feynman diagrams were originally introduced as contributions
to the flux~\cite{Mukamel_1995}. 
In this expression we substitute photon self-energy with its explicit expression, eq~S2,
separating orders of contributions to the latter.
Projections of contributions of different orders will yield analog of double-sided Feynman
diagrams corresponding to optical processes at the order of the diagram.
For example, second order double-sided Feynman diagram results from second order
contribution to $\Pi^{el}$ - first term in the right-hand-side of eq~S2:
\begin{align}
\label{Ipt2}
 I_{pt}^{(2)}(t) &= 2\,\mbox{Im}\int_{-\infty}^t dt'\, \sum_{\alpha_1,\alpha_2}
 \sum_{\begin{subarray}{c}n_1,n_2 \\ n_3,n_4\end{subarray} \in M}
U_{\alpha_1,n_1n_2}\, U_{n_3n_4,\alpha_2} 
\\ &\times
 \bigg(
 G^{<}_{n_2n_4}(t,t')\, G^{>}_{n_3n_1}(t',t)\, F^{>}_{\alpha_2,\alpha_1} (t',t)
-G^{>}_{n_2n_4}(t,t')\, G^{<}_{n_3n_1}(t',t)\, F^{<}_{\alpha_2,\alpha_1} (t',t)
 \bigg)
 \nonumber
\end{align}
The corresponding double-sided Feynman diagrams are shown in Figure~\ref{fig3}.
Two additional diagrams (accounted for by $\mbox{Im}\ldots$ in the expression above)
are obtained by switching contour branches and flipping arrows in the photon Green's function.

Similarly, fourth order double sided Feynman diagrams are obtained by
substituting fourth order contribution to self-energy $\Pi^{el}$, second term
in the right side of eq~S2, into expression for photon flux, eq~\ref{Iptdef}.
Corresponding diagrams are shown in Figure~\ref{fig4}.
Note, only projections along the contour (Green's function Hilbert space projections)
are shown. 

We note that simulating double-sided Feynman diagrams following
bare perturbation expansion is not feasible also due to the fact that 
such expansion takes into account also decoupled diagrams which 
should not contribute. Complicated subtraction of terms should be performed 
in such expansion as was discussed in Refs.~\cite{leijnse_kinetic_2008,koller_density-operator_2010}.
The problem does not appear in the present Green's function based approach~\cite{bergmann_electron_2019}.


\begin{figure}[htbp]
\centering\includegraphics[width=0.8\linewidth]{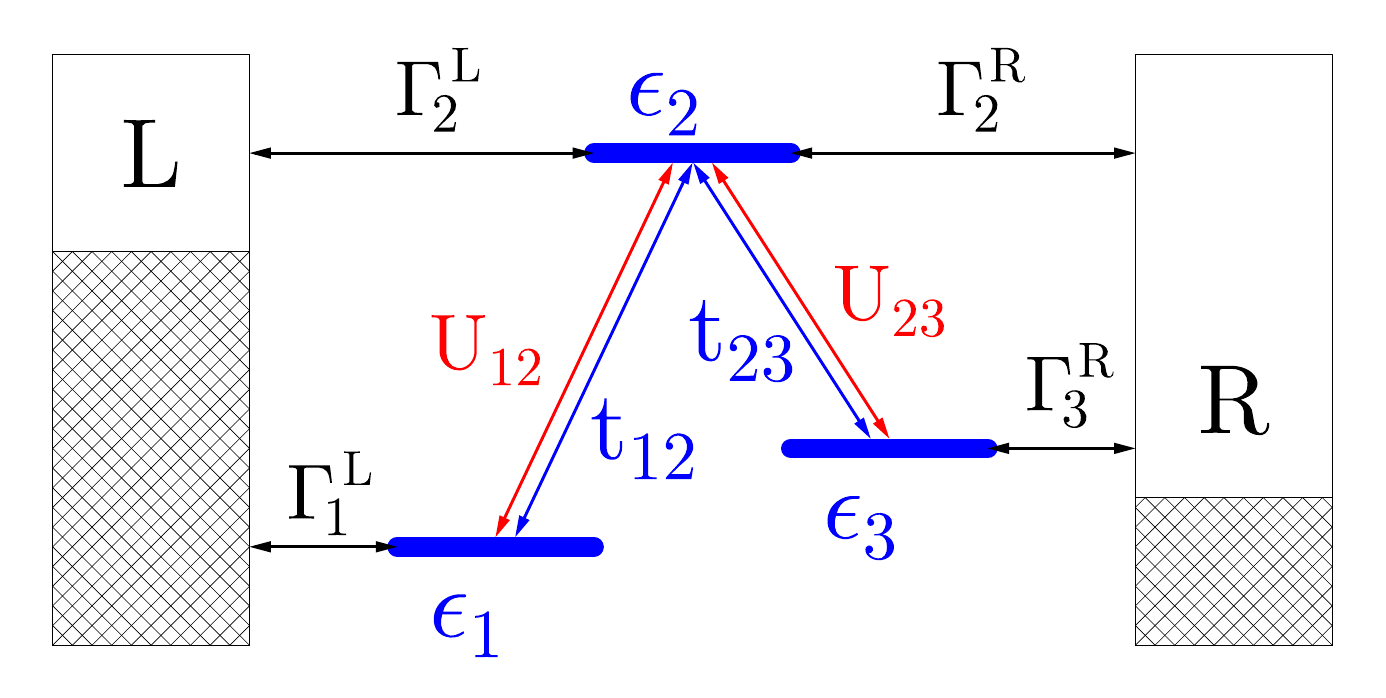}
\caption{\label{fig5}
Donor ($1$) - bridge ($2$) - acceptor ($3$) junction model for photo-assisted electron transport.
}
\end{figure}

\subsection{Numerical example}\label{numres}
The following simulations of particle and energy fluxes illustrate the conserving character of 
the double-sided Feynman diagram approach.
We assume a three level model representing donor-bridge-acceptor (DBA) molecular structure
with donor coupled to left and acceptor to right contacts. 
Bridge is assumed to be weakly coupled to both contacts (Figure~\ref{fig5}).
The donor and acceptor energies ($\varepsilon_1$ and $\varepsilon_3$)
are lower than the bridge energy ($\varepsilon_2$).
The system is subjected to external radiation which facilitates electron transfer
from donor to bridge and from bridge to acceptor (see Figure~\ref{fig5}).
The Hamiltonian is
\begin{align}
\hat H_M &= \sum_{m=1}^3 \varepsilon_m \hat d_m^\dagger\hat d_m
+ \sum_{m=1}^2\bigg( t_{m,m+1}\hat d_m^\dagger\hat d_m + H.c.\bigg)
\\
\hat V_{ML} &= \sum_{\ell in L}\bigg( V_{1\ell}\hat d_1^\dagger \hat c_\ell + V_{2\ell}\hat d_2^\dagger\hat c_\ell+H.c.\bigg)
\\
\hat V_{MR} &= \sum_{r\in R}\bigg( V_{3r}\hat d_3^\dagger\hat c_r + V_{2r}\hat d_2^\dagger\hat c_r+H.c.\bigg)
\\
\hat V_{M,rad} &= \sum_\alpha \bigg( U_{\alpha,12}\hat a_\alpha^\dagger \hat d_1^\dagger\hat d_2
+ U_{\alpha,32} \hat a_\alpha^\dagger \hat d_3^\dagger\hat d_2 + H.c.\bigg)
\end{align}
A similar model was used in Ref.~\citenum{gao_optical_2016}, where non-conserving character
of standard tools of nonlinear optical spectroscopy was illustrated.
Here we demonstrate that the present expansion satisfies conservation laws.

\begin{figure}[htbp]
\centering\includegraphics[width=0.8\linewidth]{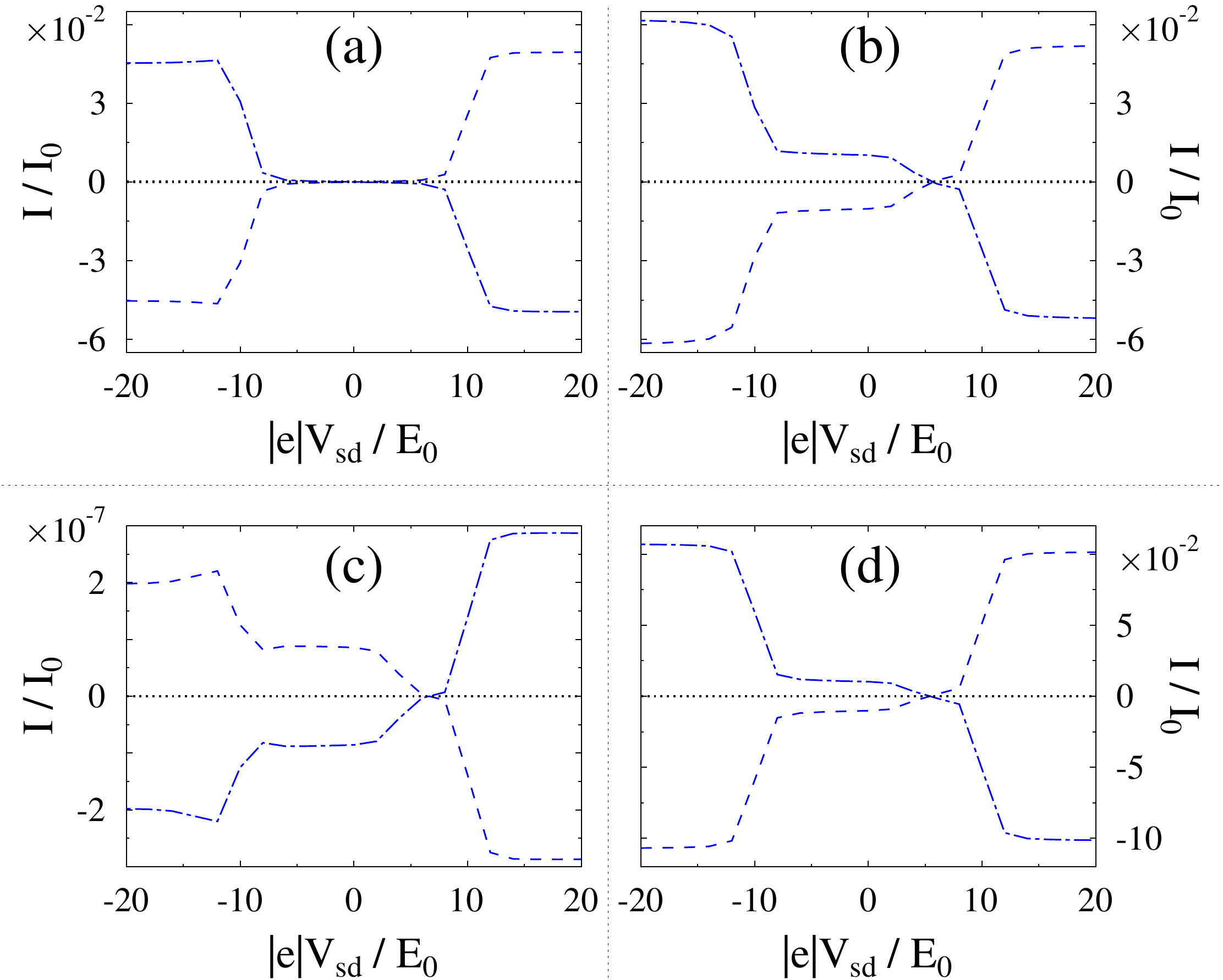}
\caption{\label{fig6}
Charge conservation, eq~\ref{charge}, for the junction model of Figure~\ref{fig5}.
Shown are $I_L$ (dashed line, blue), $I_R$ (dash-dotted line, blue) and their sum (dotted line, black)
for (a) zero, (b) second. and (c) fourth order contributions; (d) shows total fluxes, eq~\ref{IKss}.
See text for parameters.
}
\end{figure}

We focus on steady-state and check the conservation of
charge 
\begin{equation}
\label{charge}
 I_L=-I_R
\end{equation}
and energy
\begin{equation}
\label{energy}
 J_L+J_R-J_{pt}=0
\end{equation}
Note that the minus sign in the energy balance is due to opposite convention
about flux positivity for electron fluxes (positive is flux going into the system)
and photons (positive is flux going out of the system).
At steady state. all fluxes eqs~\ref{IK_NEGF}-\ref{Jpt_NEGF}, are time-independent.
They can be expressed in terms of
Fourier transforms of corresponding Green's functions and self-energies as
($K=L,R$)
\begin{align}
\label{IKss}
 I_K &= \int_{-\infty}^{+\infty}\frac{dE}{2\pi}\, i_K(E)
 \\
 \label{Iptss}
 I_{pt} &= \int_{-\infty}^{+\infty}\frac{d\omega}{2\pi}\, i_{pt}(\omega)
  \\
  \label{JKss}
  J_K &= \int_{-\infty}^{+\infty}\frac{dE}{2\pi}\, E\, i_K(E)
  \\
  \label{Jptss}
  J_{pt} &= \int_{-\infty}^{+\infty}\frac{d\omega}{2\pi}\, \omega\, i_{pt}(\omega)
\end{align}
where
\begin{align}
 i_K(E) &\equiv \mbox{Tr}\bigg[
 \Sigma_K^{<}(E)\, G^{>}(E)-\Sigma_K^{>}(E)\,G^{<}(E) \bigg]
 \\
 i_{pt}(\omega) &\equiv \mbox{Tr}\bigg[
 F^{<}(\omega)\,  \Pi^{>}(\omega) - F^{>}(\omega)\, \Pi^{<}(\omega) \bigg]
\end{align}

The radiation field is described as set of modes (oscillators).
populated by CW laser characterized by its frequency $\omega_0$,
intensity $N_0$, and bandwidth $\delta$, so that the population $N_{pt}(\omega)$ is
\begin{equation}
 N_{pt}(\omega) = N_0\frac{\delta^2}{(\omega-\omega_0)^2+\delta^2}
\end{equation}
Further details of the steady-state simulation can be found in Ref.~\citenum{gao_optical_2016}.

\begin{figure}[htbp]
\centering\includegraphics[width=0.8\linewidth]{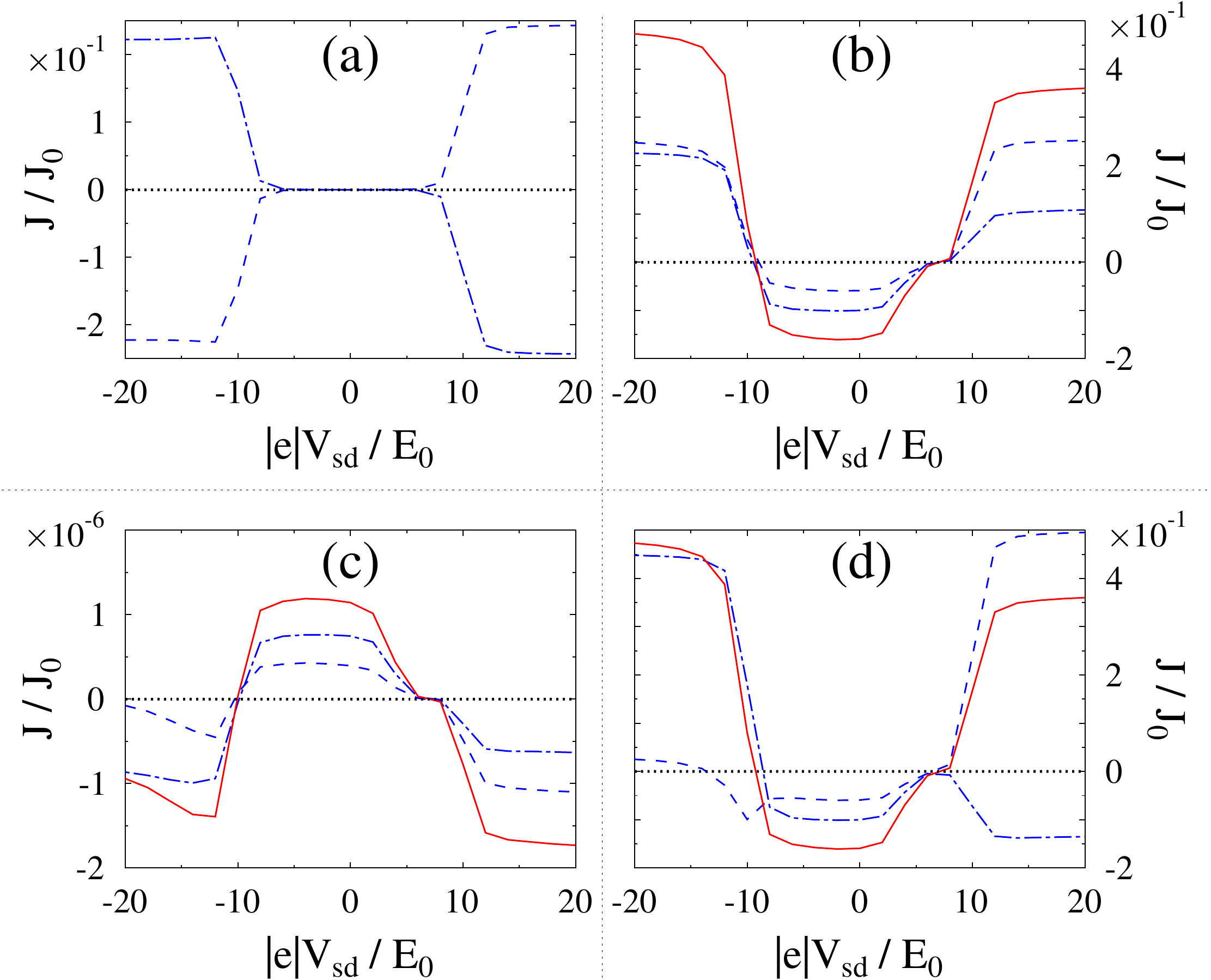}
\caption{\label{fig7}
Energy conservation, eq~\ref{energy}, for the junction model of Figure~\ref{fig5}.
Shown are $J_L$ (dashed line, blue), $J_R$ (dash-dotted line, blue),  $J_{pt}$ (solid line, red)
and their sum (dotted line, black)
for (a) zero, (b) second. and (c) fourth order contributions; (d) shows total fluxes, eqs~\ref{JKss} and \ref{Jptss}.
See text for parameters.
}
\end{figure}

The simulation parameters are (all numbers are given in terms of arbitrary unit of energy $E_0$):
$k_BT=0.25$, $\varepsilon_1=-5$, $\varepsilon_2=5$, $\varepsilon_3=-2$, $t_{12}=t_{23}=0.1$.
$\Gamma_1^L=\Gamma_3^R=1$ and $\Gamma_2^L=\Gamma_2^R=0.1$ are electron escape rates 
from donor, bridge and acceptor into left and right contacts.
$\gamma_0=0.1$ is energy escape rate from the molecule into radiation field modes. 
The molecule is subjected to external laser radiation with frequency $\omega_0=7$
and width $\delta=0.1$.
The laser frequency is chosen at resonance for the transition between bridge and acceptor.
Fermi energy is taken as the origin, $E_F=0$, and bias is assumed to be applied
symmetrically, $\mu_{L/R}=E_F\pm |e|V_{sd}/2$.
Simulations were performed on energy grid spanning region from
$-15$ to $+15$ with step $0.01$. Self-consistent NEGF simulation
was assumed to converge when levels populations difference
at consecutive steps is less than $10^{-12}$.
Results for particle and energy fluxes are presented in terms of flux units 
$I_0\equiv 1/t_0$ and $J_0\equiv E_0/t_0$, respectively
($t_0\equiv \hbar/E_0$ is unit of time).

\begin{figure}[htbp]
\centering\includegraphics[width=0.8\linewidth]{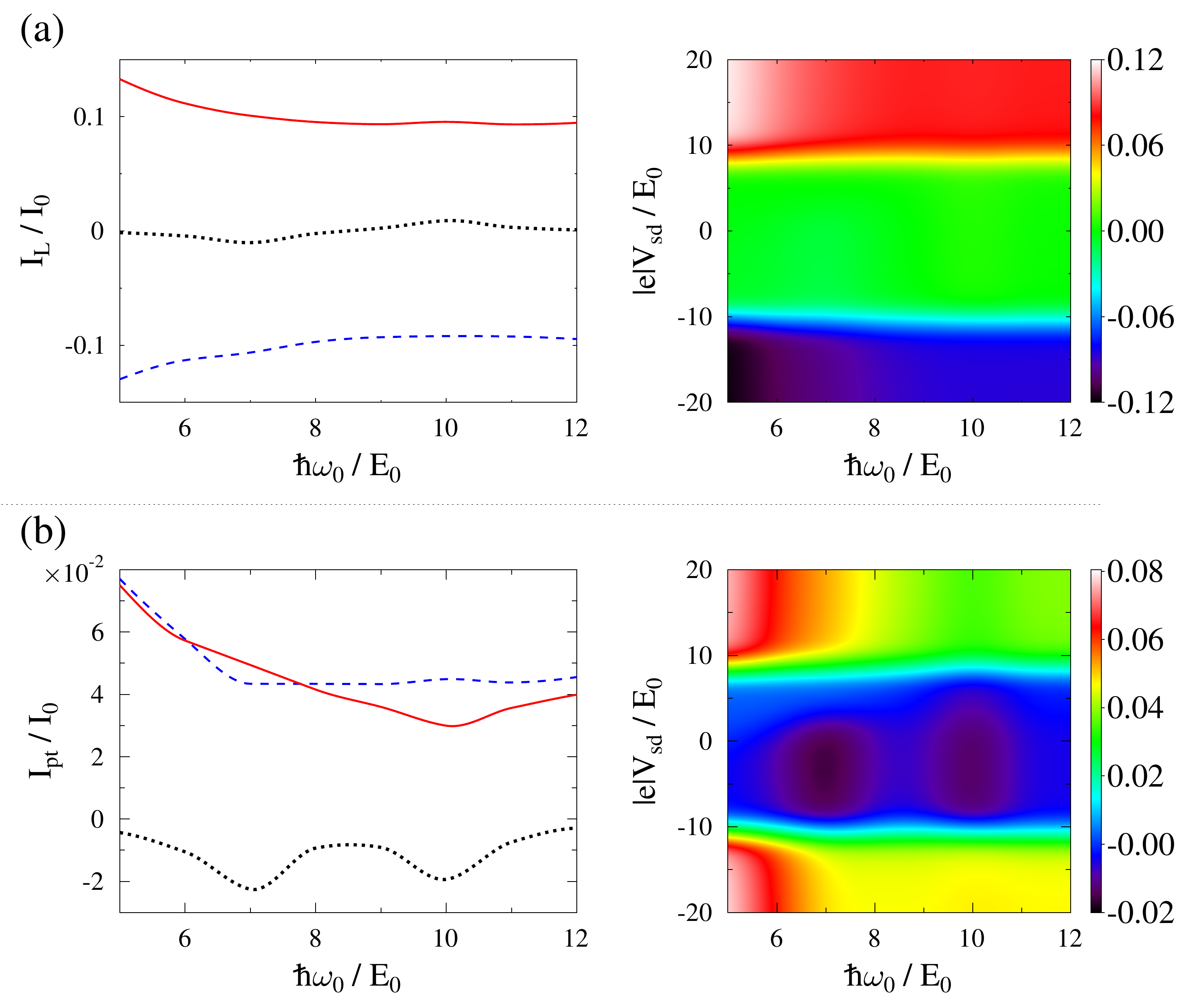}
\caption{\label{fig8}
Particle fluxes vs. pumping frequency $\omega_0$ for the junction model of Figure~\ref{fig5}.
Shown are (a) $I_L=-I_R$, eq~\ref{IKss}, and (b) $I_{pt}$, eq~\ref{Iptss},
at biases $|e|\, V_{sd}=-16\, E_0$ (dashed line, blue), $|e|\, V_{sd}=0$ (dotted line, black), 
and $|e|\, V_{sd}=16\, E_0$ (solid line, red) in left panels.
Right column shows map of the fluxes vs. pumping frequency $\omega_0$ and bias $V_{sd}$.
See text for parameters.
}
\end{figure}

\begin{figure}[htbp]
\centering\includegraphics[width=0.8\linewidth]{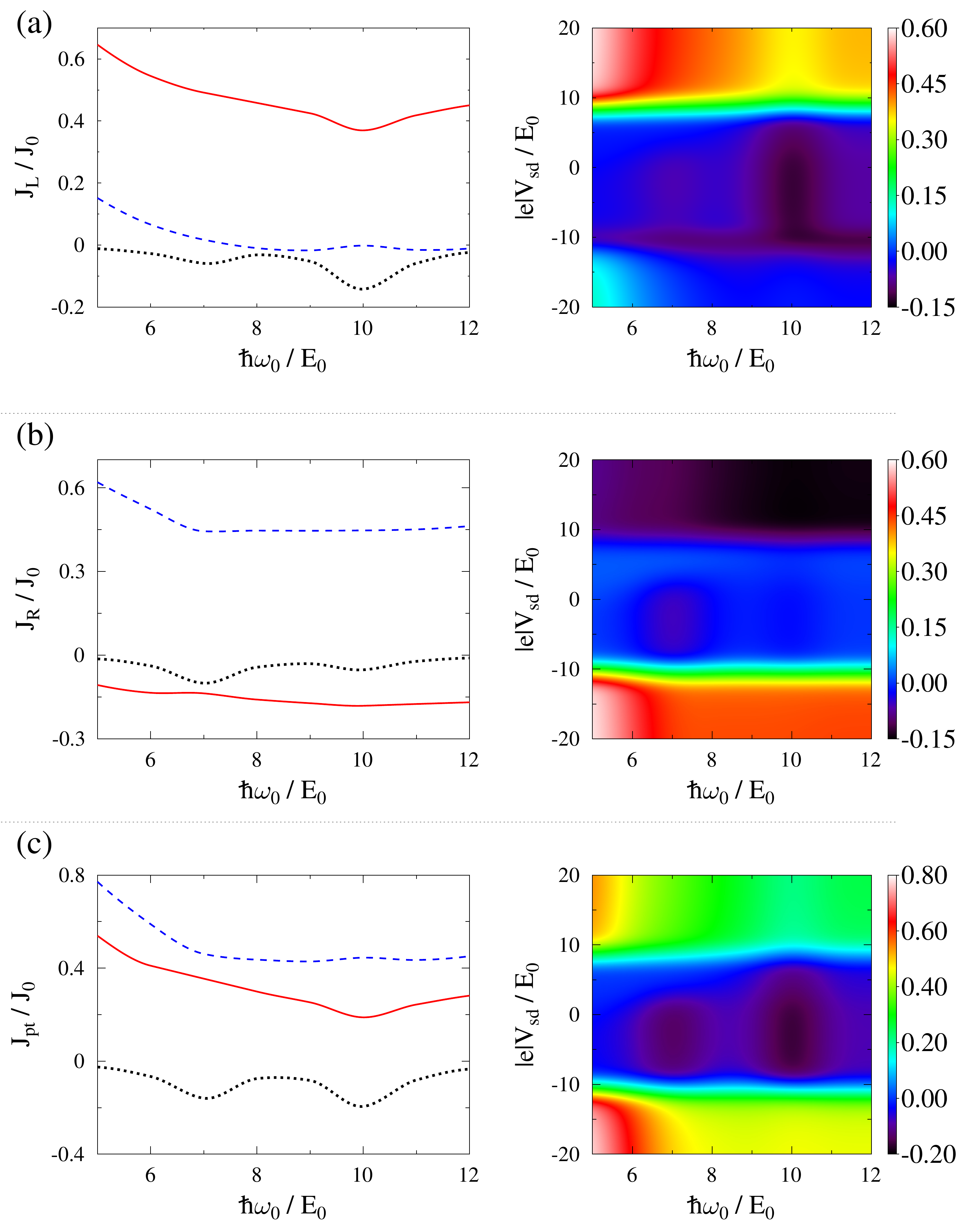}
\caption{\label{fig9}
Energy fluxes vs. pumping frequency $\omega_0$ for the junction model of Figure~\ref{fig5}.
Shown are (a) $J_L$, eq~\ref{JKss}, (b) $J_R$, eq~\ref{JKss}, and (c) $J_{pt}$, eq~\ref{Jptss}, 
at biases $|e|\, V_{sd}=-16\, E_0$ (dashed line, blue), $|e|\, V_{sd}=0$ (dotted line, black), 
and $|e|\, V_{sd}=16\, E_0$ (solid line, red) in left panels.
Right column shows map of the fluxes vs. pumping frequency $\omega_0$ and bias $V_{sd}$.
See text for parameters.
}
\end{figure}

Figure~\ref{fig6} shows charge currents, eq~\ref{IKss}, at the left and right interfaces (dashed and dash=dotted lines, respectively).
Their sum (dotted line) by charge conservation, eq~\ref{charge}, should be zero at steady-state.
Panels (a)-(c) present contributions to the fluxes of the zero, second, and fourth order diagrams in 
molecule-radiation field coupling strength; panel (d) shows sum of all the contributions.
Similarly, Figure~\ref{fig7} shows energy currents due to electrons, eq~\ref{JKss} at the left (dashed line) and the right (dash-dotted line)
interfaces and due to photons (solid line), eq~\ref{Jptss}. Their sum (dotted line) by energy conservation, eq~\ref{energy},
is zero at steady-state.
Note that the conservation laws are satisfied at each order of our diagrammatic expansion in light matter interaction, 
i.e. the sum of all double sided Feynman diagrams of a particular order satisfies charge and energy conservation.

Figures~\ref{fig8} and \ref{fig9} present spectroscopy of particle, eqs~\ref{IKss} and \ref{Iptss}, 
and energy, eqs~\ref{JKss} and \ref{Jptss}, fluxes for the junction model of Figure~\ref{fig5}. 
While for the choice of parameters charge current (Figure~\ref{fig8}a) mostly depends on bias,
the photon flux (Figure~\ref{fig8}b) is sensitive to the radiation field frequency.
At zero bias the photon flux has dips at molecular resonances
$\varepsilon_2-\varepsilon_1=10\, E_0$ and $\varepsilon_2-\varepsilon_3=7\, E_0$
due to photon absorption by the electronic system (see dotted line and map in Fig.~\ref{fig8}b). 
At higher biases laser induced absorption competes with bias induced emission.
Thus, the photon flux is suppressed at molecular resonances (see solid an dashed line in Fig.~\ref{fig8}b).

The energy fluxes show a similar frequency dependence.
In particular, dips in $J_L$, $J_R$ and $J_{pt}$ 
at donor-bridge molecular resonance  $\varepsilon_2-\varepsilon_1=10$ 
and $\varepsilon_2-\varepsilon_1=10$at low biases
(see dotted lines  and maps in Figures~\ref{fig9}a, b and c) 
indicate increased $L$ to $M$ and $R$ to $M$ energy fluxes caused by increased electron transfer into the donor and acceptor 
facilitated by the radiation field pumping.
Note that for $\varepsilon_1=-5\, E_0$ and $\varepsilon_3=-2\, E_0$ 
energy flux coming from the contacts will be negative.
Similarly, dips in the photon flux indicate increase in energy coming into the system.
At higher biases radiation field pumping is counteracted by bias induced emission,
so that the energy curves become smoother, although $J_L$ and $J_{pt}$ still show dips at 
molecular resonance corresponding to donor-bridge transition (see solid lines in Figures~\ref{fig9}a and c). 

Note that while in the numerical illustration we focus on steady-state,
where the initial state of the field and way of switching on of the light-matter interaction are not important,
description of light pulses will require solving corresponding time-dependent problem, eqs~\ref{IK_NEGF}-\ref{Jpt_NEGF}


\section{Conclusions}\label{conclude}
We had developed a theoretical description of optical spectroscopy for open nonequilibrium systems,
where both molecular degrees of freedom and radiation field are treated quantum 
mechanically and where charge and energy conservations in the system are built in.
Starting from nonequilibrium Green's function formulations we 
show connection with Liouville space description and 
introduce generalization of  double-sided Feynman diagrams.
The latter is standard tool widely used by theorists
and experimentalists for design and interpretation of experiments.

We performed an expansion in the light-matter coupling strength
within the standard NEGF, and presented different contributions to the photon flux by
double-sided Feynman diagrams. In particular, the order of diagrammatic expansion
in photon self-energy due to coupling to electrons is identified as 
order of optical process. Double-sided Feynman diagrams
are shown to be projections of corresponding Feynman diagrams on
the Keldysh contour. Light-matter interaction events in double-sided Feynman diagrams 
are accompanied by change of molecular orbital, as is expected for
weak coupling case.

Our study  bridges the theoretical approaches used in quantum transport and  optical spectroscopy. 
It establishes firm theoretical basis for applying traditional tools
of nonlinear optical spectroscopy in molecular optoelectronics.
Developing theoretical description of optical spectroscopy for strongly interacting open molecular systems is a goal for future research.

\begin{suppinfo}
Supporting information. 
Explicit expressions for the self-energies of electrons due to coupling to radiation field modes and photons due 
to coupling to electrons to fourth order in light-matter interaction.
\end{suppinfo}


\begin{acknowledgement}
This material is based upon work supported by the National Science Foundation under
Grants No. CHE-1565939 (M.G.) and No. CHE-1663822 (S.M.).
\end{acknowledgement}


\begin{mcitethebibliography}{55}
\providecommand*\natexlab[1]{#1}
\providecommand*\mciteSetBstSublistMode[1]{}
\providecommand*\mciteSetBstMaxWidthForm[2]{}
\providecommand*\mciteBstWouldAddEndPuncttrue
  {\def\EndOfBibitem{\unskip.}}
\providecommand*\mciteBstWouldAddEndPunctfalse
  {\let\EndOfBibitem\relax}
\providecommand*\mciteSetBstMidEndSepPunct[3]{}
\providecommand*\mciteSetBstSublistLabelBeginEnd[3]{}
\providecommand*\EndOfBibitem{}
\mciteSetBstSublistMode{f}
\mciteSetBstMaxWidthForm{subitem}{(\alph{mcitesubitemcount})}
\mciteSetBstSublistLabelBeginEnd
  {\mcitemaxwidthsubitemform\space}
  {\relax}
  {\relax}

\bibitem[Krausz and Ivanov(2009)Krausz, and Ivanov]{krausz_attosecond_2009}
Krausz,~F.; Ivanov,~M. Attosecond Physics. \emph{Rev. Mod. Phys.}
  \textbf{2009}, \emph{81}, 163--234\relax
\mciteBstWouldAddEndPuncttrue
\mciteSetBstMidEndSepPunct{\mcitedefaultmidpunct}
{\mcitedefaultendpunct}{\mcitedefaultseppunct}\relax
\EndOfBibitem
\bibitem[Kawata \latin{et~al.}(2009)Kawata, Inouye, and
  Verma]{kawata_plasmonics_2009}
Kawata,~S.; Inouye,~Y.; Verma,~P. Plasmonics for Near-Field Nano-Imaging and
  Superlensing. \emph{Nature Photon.} \textbf{2009}, \emph{3}, 388--394\relax
\mciteBstWouldAddEndPuncttrue
\mciteSetBstMidEndSepPunct{\mcitedefaultmidpunct}
{\mcitedefaultendpunct}{\mcitedefaultseppunct}\relax
\EndOfBibitem
\bibitem[Le~Ru and Etchegoin(2012)Le~Ru, and
  Etchegoin]{le_ru_single-molecule_2012}
Le~Ru,~E.~C.; Etchegoin,~P.~G. Single-{Molecule} {Surface}-{Enhanced} {Raman}
  {Spectroscopy}. \emph{Ann. Rev. Phys. Chem.} \textbf{2012}, \emph{63},
  65--87\relax
\mciteBstWouldAddEndPuncttrue
\mciteSetBstMidEndSepPunct{\mcitedefaultmidpunct}
{\mcitedefaultendpunct}{\mcitedefaultseppunct}\relax
\EndOfBibitem
\bibitem[Verma(2017)]{verma_tip-enhanced_2017}
Verma,~P. Tip-{Enhanced} {Raman} {Spectroscopy}: {Technique} and {Recent}
  {Advances}. \emph{Chem. Rev.} \textbf{2017}, \emph{117}, 6447--6466\relax
\mciteBstWouldAddEndPuncttrue
\mciteSetBstMidEndSepPunct{\mcitedefaultmidpunct}
{\mcitedefaultendpunct}{\mcitedefaultseppunct}\relax
\EndOfBibitem
\bibitem[Kampfrath \latin{et~al.}(2013)Kampfrath, Tanaka, and
  Nelson]{kampfrath_resonant_2013}
Kampfrath,~T.; Tanaka,~K.; Nelson,~K.~A. Resonant and Nonresonant Control Over
  Matter and Light by Intense Terahertz Transients. \emph{Nature Photon.}
  \textbf{2013}, \emph{7}, 680--690\relax
\mciteBstWouldAddEndPuncttrue
\mciteSetBstMidEndSepPunct{\mcitedefaultmidpunct}
{\mcitedefaultendpunct}{\mcitedefaultseppunct}\relax
\EndOfBibitem
\bibitem[Young \latin{et~al.}(2018)Young, Ueda, G{\" u}hr, Bucksbaum, Simon,
  Mukamel, Rohringer, Prince, Masciovecchio, Meyer, Rudenko, Rolles, Bostedt,
  Fuchs, Reis, Santra, Kapteyn, Murnane, Ibrahim, L{\' e}gar{\' e}, Vrakking,
  Isinger, Kroon, Gisselbrecht, L'Huillier, W{\" o}rner, and
  Leone]{young_roadmap_2018}
Young,~L.; Ueda,~K.; G{\" u}hr,~M.; Bucksbaum,~P.~H.; Simon,~M.; Mukamel,~S.;
  Rohringer,~N.; Prince,~K.~C.; Masciovecchio,~C.; Meyer,~M. \latin{et~al.}
  Roadmap of Ultrafast X-ray Atomic and Molecular Physics. \emph{J. Phys. B:
  At., Mol. Opt. Phys.} \textbf{2018}, \emph{51}, 032003\relax
\mciteBstWouldAddEndPuncttrue
\mciteSetBstMidEndSepPunct{\mcitedefaultmidpunct}
{\mcitedefaultendpunct}{\mcitedefaultseppunct}\relax
\EndOfBibitem
\bibitem[Bennett \latin{et~al.}(2018)Bennett, Kowalewski, Rouxel, and
  Mukamel]{bennett_monitoring_2018}
Bennett,~K.; Kowalewski,~M.; Rouxel,~J.~R.; Mukamel,~S. Monitoring Molecular
  Nonadiabatic Dynamics with Femtosecond {X}-ray Diffraction. \emph{Proc. Natl.
  Acad. Sci.} \textbf{2018}, \emph{115}, 6538--6547\relax
\mciteBstWouldAddEndPuncttrue
\mciteSetBstMidEndSepPunct{\mcitedefaultmidpunct}
{\mcitedefaultendpunct}{\mcitedefaultseppunct}\relax
\EndOfBibitem
\bibitem[Schlawin \latin{et~al.}(2018)Schlawin, Dorfman, and
  Mukamel]{schlawin_entangled_2018}
Schlawin,~F.; Dorfman,~K.~E.; Mukamel,~S. Entangled {Two}-{Photon} {Absorption}
  {Spectroscopy}. \emph{Acc. Chem. Res.} \textbf{2018}, \emph{51},
  2207--2214\relax
\mciteBstWouldAddEndPuncttrue
\mciteSetBstMidEndSepPunct{\mcitedefaultmidpunct}
{\mcitedefaultendpunct}{\mcitedefaultseppunct}\relax
\EndOfBibitem
\bibitem[Dorfman \latin{et~al.}(2019)Dorfman, Asban, Ye, Rouxel, Cho, and
  Mukamel]{dorfman_monitoring_2019}
Dorfman,~K.~E.; Asban,~S.; Ye,~L.; Rouxel,~J.~R.; Cho,~D.; Mukamel,~S.
  Monitoring {Spontaneous} {Charge}-{Density} {Fluctuations} by
  {Single}-{Molecule} {Diffraction} of {Quantum} {Light}. \emph{J. Phys. Chem.
  Lett.} \textbf{2019}, \emph{10}, 768--773\relax
\mciteBstWouldAddEndPuncttrue
\mciteSetBstMidEndSepPunct{\mcitedefaultmidpunct}
{\mcitedefaultendpunct}{\mcitedefaultseppunct}\relax
\EndOfBibitem
\bibitem[Asban \latin{et~al.}(2019)Asban, Dorfman, and
  Mukamel]{asban_quantum_2019}
Asban,~S.; Dorfman,~K.~E.; Mukamel,~S. Quantum Phase-Sensitive Diffraction and
  Imaging Using Entangled Photons. \emph{Proc. Natl. Acad. Sci.} \textbf{2019},
  201904839\relax
\mciteBstWouldAddEndPuncttrue
\mciteSetBstMidEndSepPunct{\mcitedefaultmidpunct}
{\mcitedefaultendpunct}{\mcitedefaultseppunct}\relax
\EndOfBibitem
\bibitem[Qiu \latin{et~al.}(2003)Qiu, Nazin, and Ho]{qiu_vibrationally_2003}
Qiu,~X.~H.; Nazin,~G.~V.; Ho,~W. Vibrationally {Resolved} {Fluorescence}
  {Excited} with {Submolecular} {Precision}. \emph{Science} \textbf{2003},
  \emph{299}, 542--546\relax
\mciteBstWouldAddEndPuncttrue
\mciteSetBstMidEndSepPunct{\mcitedefaultmidpunct}
{\mcitedefaultendpunct}{\mcitedefaultseppunct}\relax
\EndOfBibitem
\bibitem[Dong \latin{et~al.}(2004)Dong, Guo, Trifonov, Dorozhkin, Miki, Kimura,
  Yokoyama, and Mashiko]{dong_vibrationally_2004}
Dong,~Z.-C.; Guo,~X.-L.; Trifonov,~A.~S.; Dorozhkin,~P.~S.; Miki,~K.;
  Kimura,~K.; Yokoyama,~S.; Mashiko,~S. Vibrationally {Resolved} {Fluorescence}
  from {Organic} {Molecules} near {Metal} {Surfaces} in a {Scanning}
  {Tunneling} {Microscope}. \emph{Phys. Rev. Lett.} \textbf{2004}, \emph{92},
  086801\relax
\mciteBstWouldAddEndPuncttrue
\mciteSetBstMidEndSepPunct{\mcitedefaultmidpunct}
{\mcitedefaultendpunct}{\mcitedefaultseppunct}\relax
\EndOfBibitem
\bibitem[Wu \latin{et~al.}(2008)Wu, Nazin, and Ho]{HoPRB08}
Wu,~S.~W.; Nazin,~G.~V.; Ho,~W. Intramolecular Photon Emission from a Single
  Molecule in a Scanning Tunneling Microscope. \emph{Phys. Rev. B}
  \textbf{2008}, \emph{77}, 205430\relax
\mciteBstWouldAddEndPuncttrue
\mciteSetBstMidEndSepPunct{\mcitedefaultmidpunct}
{\mcitedefaultendpunct}{\mcitedefaultseppunct}\relax
\EndOfBibitem
\bibitem[Chen \latin{et~al.}(2010)Chen, Chu, Bobisch, Mills, and
  Ho]{chen_viewing_2010}
Chen,~C.; Chu,~P.; Bobisch,~C.~A.; Mills,~D.~L.; Ho,~W. Viewing the {Interior}
  of a {Single} {Molecule}: {Vibronically} {Resolved} {Photon} {Imaging} at
  {Submolecular} {Resolution}. \emph{Phys. Rev. Lett.} \textbf{2010},
  \emph{105}, 217402\relax
\mciteBstWouldAddEndPuncttrue
\mciteSetBstMidEndSepPunct{\mcitedefaultmidpunct}
{\mcitedefaultendpunct}{\mcitedefaultseppunct}\relax
\EndOfBibitem
\bibitem[Zhang \latin{et~al.}(2016)Zhang, Luo, Zhang, Yu, Kuang, Zhang, Meng,
  Luo, Yang, Dong, and Hou]{zhang_visualizing_2016}
Zhang,~Y.; Luo,~Y.; Zhang,~Y.; Yu,~Y.-J.; Kuang,~Y.-M.; Zhang,~L.; Meng,~Q.-S.;
  Luo,~Y.; Yang,~J.-L.; Dong,~Z.-C. \latin{et~al.}  Visualizing Coherent
  Intermolecular Dipole-Dipole Coupling in Real Space. \emph{Nature}
  \textbf{2016}, \emph{531}, 623--627\relax
\mciteBstWouldAddEndPuncttrue
\mciteSetBstMidEndSepPunct{\mcitedefaultmidpunct}
{\mcitedefaultendpunct}{\mcitedefaultseppunct}\relax
\EndOfBibitem
\bibitem[Imada \latin{et~al.}(2016)Imada, Miwa, Imai-Imada, Kawahara, Kimura,
  and Kim]{imada_real-space_2016}
Imada,~H.; Miwa,~K.; Imai-Imada,~M.; Kawahara,~S.; Kimura,~K.; Kim,~Y.
  Real-Space Investigation of Energy Transfer in Heterogeneous Molecular
  Dimers. \emph{Nature} \textbf{2016}, \emph{538}, 364--367\relax
\mciteBstWouldAddEndPuncttrue
\mciteSetBstMidEndSepPunct{\mcitedefaultmidpunct}
{\mcitedefaultendpunct}{\mcitedefaultseppunct}\relax
\EndOfBibitem
\bibitem[Kimura \latin{et~al.}(2019)Kimura, Miwa, Imada, Imai-Imada, Kawahara,
  Takeya, Kawai, Galperin, and Kim]{kimura_selective_2019}
Kimura,~K.; Miwa,~K.; Imada,~H.; Imai-Imada,~M.; Kawahara,~S.; Takeya,~J.;
  Kawai,~M.; Galperin,~M.; Kim,~Y. Selective Triplet Exciton Formation in a
  Single Molecule. \emph{Nature} \textbf{2019}, \emph{570}, 210--213\relax
\mciteBstWouldAddEndPuncttrue
\mciteSetBstMidEndSepPunct{\mcitedefaultmidpunct}
{\mcitedefaultendpunct}{\mcitedefaultseppunct}\relax
\EndOfBibitem
\bibitem[Schneider \latin{et~al.}(2012)Schneider, L\"u, Brandbyge, and
  Berndt]{BerndtPRL12}
Schneider,~N.~L.; L\"u,~J.~T.; Brandbyge,~M.; Berndt,~R. Light Emission Probing
  Quantum Shot Noise and Charge Fluctuations at a Biased Molecular Junction.
  \emph{Phys. Rev. Lett.} \textbf{2012}, \emph{109}, 186601\relax
\mciteBstWouldAddEndPuncttrue
\mciteSetBstMidEndSepPunct{\mcitedefaultmidpunct}
{\mcitedefaultendpunct}{\mcitedefaultseppunct}\relax
\EndOfBibitem
\bibitem[Natelson \latin{et~al.}(2013)Natelson, Li, and
  Herzog]{natelson_nanogap_2013}
Natelson,~D.; Li,~Y.; Herzog,~J.~B. Nanogap Structures: Combining Enhanced
  {Raman} Spectroscopy and Electronic Transport. \emph{Phys. Chem. Chem. Phys.}
  \textbf{2013}, \emph{15}, 5262--5275\relax
\mciteBstWouldAddEndPuncttrue
\mciteSetBstMidEndSepPunct{\mcitedefaultmidpunct}
{\mcitedefaultendpunct}{\mcitedefaultseppunct}\relax
\EndOfBibitem
\bibitem[Ward \latin{et~al.}(2008)Ward, Halas, Ciszek, Tour, Wu, Nordlander,
  and Natelson]{NatelsonNL08}
Ward,~D.~R.; Halas,~N.~J.; Ciszek,~J.~W.; Tour,~J.~M.; Wu,~Y.; Nordlander,~P.;
  Natelson,~D. Simultaneous Measurements of Electronic Conduction and Raman
  Response in Molecular Junctions. \emph{Nano Lett.} \textbf{2008}, \emph{8},
  919--924\relax
\mciteBstWouldAddEndPuncttrue
\mciteSetBstMidEndSepPunct{\mcitedefaultmidpunct}
{\mcitedefaultendpunct}{\mcitedefaultseppunct}\relax
\EndOfBibitem
\bibitem[Ioffe \latin{et~al.}(2008)Ioffe, Shamai, Ophir, Noy, Yutsis, Kfir,
  Cheshnovsky, and Selzer]{CheshnovskySelzerNatNano08}
Ioffe,~Z.; Shamai,~T.; Ophir,~A.; Noy,~G.; Yutsis,~I.; Kfir,~K.;
  Cheshnovsky,~O.; Selzer,~Y. Detection of Heating in Current-Carrying
  Molecular Junctions by Raman Scattering. \emph{Nature Nanotech.}
  \textbf{2008}, \emph{3}, 727--732\relax
\mciteBstWouldAddEndPuncttrue
\mciteSetBstMidEndSepPunct{\mcitedefaultmidpunct}
{\mcitedefaultendpunct}{\mcitedefaultseppunct}\relax
\EndOfBibitem
\bibitem[Ward \latin{et~al.}(2011)Ward, Corley, Tour, and
  Natelson]{NatelsonNatNano11}
Ward,~D.~R.; Corley,~D.~A.; Tour,~J.~M.; Natelson,~D. Vibrational and
  Electronic Heating in Nanoscale Junctions. \emph{Nature Nanotech.}
  \textbf{2011}, \emph{6}, 33--38\relax
\mciteBstWouldAddEndPuncttrue
\mciteSetBstMidEndSepPunct{\mcitedefaultmidpunct}
{\mcitedefaultendpunct}{\mcitedefaultseppunct}\relax
\EndOfBibitem
\bibitem[Grosse \latin{et~al.}(2013)Grosse, Etzkorn, Kuhnke, Loth, and
  Kern]{LothAPL13}
Grosse,~C.; Etzkorn,~M.; Kuhnke,~K.; Loth,~S.; Kern,~K. Quantitative Mapping of
  Fast Voltage Pulses in Tunnel Junctions by Plasmonic Luminescence.
  \emph{Appl. Phys. Lett.} \textbf{2013}, \emph{103}, 183108\relax
\mciteBstWouldAddEndPuncttrue
\mciteSetBstMidEndSepPunct{\mcitedefaultmidpunct}
{\mcitedefaultendpunct}{\mcitedefaultseppunct}\relax
\EndOfBibitem
\bibitem[Galperin and Nitzan(2012)Galperin, and Nitzan]{MGANPCCP12}
Galperin,~M.; Nitzan,~A. Molecular Optoelectronics: The Interaction of
  Molecular Conduction Junctions with Light. \emph{Phys. Chem. Chem. Phys.}
  \textbf{2012}, \emph{14}, 9421--9438\relax
\mciteBstWouldAddEndPuncttrue
\mciteSetBstMidEndSepPunct{\mcitedefaultmidpunct}
{\mcitedefaultendpunct}{\mcitedefaultseppunct}\relax
\EndOfBibitem
\bibitem[Okumura and Tanimura(1997)Okumura, and Tanimura]{OkumuraJCP97}
Okumura,~K.; Tanimura,~Y. The (2n+1)th-order Off-Resonant Spectroscopy from the
  (n+1)th-order Anharmonicities of Molecular Vibrational Modes in the Condensed
  Phase. \emph{J. Chem. Phys.} \textbf{1997}, \emph{106}, 1687--1698\relax
\mciteBstWouldAddEndPuncttrue
\mciteSetBstMidEndSepPunct{\mcitedefaultmidpunct}
{\mcitedefaultendpunct}{\mcitedefaultseppunct}\relax
\EndOfBibitem
\bibitem[Okumura and Tanimura(1997)Okumura, and Tanimura]{OkumuraTanimuraJCP97}
Okumura,~K.; Tanimura,~Y. Femtosecond Two-Dimensional Spectroscopy from
  Anharmonic Vibrational Modes of Molecules in the Condensed Phase. \emph{J.
  Chem. Phys.} \textbf{1997}, \emph{107}, 2267--2283\relax
\mciteBstWouldAddEndPuncttrue
\mciteSetBstMidEndSepPunct{\mcitedefaultmidpunct}
{\mcitedefaultendpunct}{\mcitedefaultseppunct}\relax
\EndOfBibitem
\bibitem[Xu \latin{et~al.}(2001)Xu, , and Fleming]{FlemingJPCA01}
Xu,~Q.-H.; ; Fleming,~G.~R. Isomerization Dynamics of
  1,1’-Diethyl-4,4’-Cyanine (1144C) Studied by Different Third-Order
  Nonlinear Spectroscopic Measurements. \emph{J. Phys. Chem. A} \textbf{2001},
  \emph{105}, 10187--10195\relax
\mciteBstWouldAddEndPuncttrue
\mciteSetBstMidEndSepPunct{\mcitedefaultmidpunct}
{\mcitedefaultendpunct}{\mcitedefaultseppunct}\relax
\EndOfBibitem
\bibitem[Ovchinnikov \latin{et~al.}(2001)Ovchinnikov, Apkarian, and
  Voth]{OvchinnikovApkarianVothJCP01}
Ovchinnikov,~M.; Apkarian,~V.~A.; Voth,~G.~A. Semiclassical Molecular Dynamics
  Computation of Spontaneous Light Emission in the Condensed Phase: Resonance
  Raman Spectra. \emph{J. Chem. Phys.} \textbf{2001}, \emph{114},
  7130--7143\relax
\mciteBstWouldAddEndPuncttrue
\mciteSetBstMidEndSepPunct{\mcitedefaultmidpunct}
{\mcitedefaultendpunct}{\mcitedefaultseppunct}\relax
\EndOfBibitem
\bibitem[Okumura and Tanimura(2003)Okumura, and Tanimura]{OkumuraJPCA03}
Okumura,~K.; Tanimura,~Y. Energy-Level Diagrams and Their Contribution to
  Fifth-Order Raman and Second-Order Infrared Responses:  Distinction between
  Relaxation Models by Two-Dimensional Spectroscopy†. \emph{J. Phys. Chem. A}
  \textbf{2003}, \emph{107}, 8092--8105\relax
\mciteBstWouldAddEndPuncttrue
\mciteSetBstMidEndSepPunct{\mcitedefaultmidpunct}
{\mcitedefaultendpunct}{\mcitedefaultseppunct}\relax
\EndOfBibitem
\bibitem[Mukamel \latin{et~al.}(2004)Mukamel, , and
  Abramavicius]{MukamelChemRev04}
Mukamel,~S.; ; Abramavicius,~D. Many-Body Approaches for Simulating Coherent
  Nonlinear Spectroscopies of Electronic and Vibrational Excitons. \emph{Chem.
  Rev.} \textbf{2004}, \emph{104}, 2073--2098\relax
\mciteBstWouldAddEndPuncttrue
\mciteSetBstMidEndSepPunct{\mcitedefaultmidpunct}
{\mcitedefaultendpunct}{\mcitedefaultseppunct}\relax
\EndOfBibitem
\bibitem[{\v S}anda and Mukamel(2005){\v S}anda, and Mukamel]{MukamelPRA05}
{\v S}anda,~F.; Mukamel,~S. Liouville-Space Pathways for Spectral Diffusion in
  Photon Statistics from Single Molecules. \emph{Phys. Rev. A} \textbf{2005},
  \emph{71}, 033807\relax
\mciteBstWouldAddEndPuncttrue
\mciteSetBstMidEndSepPunct{\mcitedefaultmidpunct}
{\mcitedefaultendpunct}{\mcitedefaultseppunct}\relax
\EndOfBibitem
\bibitem[Yang and Mukamel(2008)Yang, and Mukamel]{MukamelPRB08}
Yang,~L.; Mukamel,~S. Revealing Exciton-Exciton Couplings in Semiconductors
  Using Multidimensional Four-Wave Mixing Signals. \emph{Phys. Rev. B}
  \textbf{2008}, \emph{77}, 075335\relax
\mciteBstWouldAddEndPuncttrue
\mciteSetBstMidEndSepPunct{\mcitedefaultmidpunct}
{\mcitedefaultendpunct}{\mcitedefaultseppunct}\relax
\EndOfBibitem
\bibitem[Harbola and Mukamel(2009)Harbola, and Mukamel]{MukamelPRB09}
Harbola,~U.; Mukamel,~S. Coherent Stimulated X-ray Raman Spectroscopy:
  Attosecond Extension of Resonant Inelastic X-ray Raman Scattering.
  \emph{Phys. Rev. B} \textbf{2009}, \emph{79}, 085108\relax
\mciteBstWouldAddEndPuncttrue
\mciteSetBstMidEndSepPunct{\mcitedefaultmidpunct}
{\mcitedefaultendpunct}{\mcitedefaultseppunct}\relax
\EndOfBibitem
\bibitem[Mukamel(1995)]{Mukamel_1995}
Mukamel,~S. \emph{Principles of Nonlinear Optical Spectroscopy}; Oxford
  University Press, 1995\relax
\mciteBstWouldAddEndPuncttrue
\mciteSetBstMidEndSepPunct{\mcitedefaultmidpunct}
{\mcitedefaultendpunct}{\mcitedefaultseppunct}\relax
\EndOfBibitem
\bibitem[Breuer and Petruccione(2003)Breuer, and
  Petruccione]{breuer_theory_2003}
Breuer,~H.-P.; Petruccione,~F. \emph{The {Theory} of {Open} {Quantum}
  {Systems}}; Oxford University Press, 2003\relax
\mciteBstWouldAddEndPuncttrue
\mciteSetBstMidEndSepPunct{\mcitedefaultmidpunct}
{\mcitedefaultendpunct}{\mcitedefaultseppunct}\relax
\EndOfBibitem
\bibitem[Gao and Galperin(2016)Gao, and Galperin]{gao_simulation_2016}
Gao,~Y.; Galperin,~M. Simulation of Optical Response Functions in Molecular
  Junctions. \emph{J. Chem. Phys.} \textbf{2016}, \emph{144}, 244106\relax
\mciteBstWouldAddEndPuncttrue
\mciteSetBstMidEndSepPunct{\mcitedefaultmidpunct}
{\mcitedefaultendpunct}{\mcitedefaultseppunct}\relax
\EndOfBibitem
\bibitem[Galperin(2017)]{galperin_photonics_2017}
Galperin,~M. Photonics and Spectroscopy in Nanojunctions: A Theoretical
  Insight. \emph{Chem. Soc. Rev.} \textbf{2017}, \emph{46}, 4000--4019\relax
\mciteBstWouldAddEndPuncttrue
\mciteSetBstMidEndSepPunct{\mcitedefaultmidpunct}
{\mcitedefaultendpunct}{\mcitedefaultseppunct}\relax
\EndOfBibitem
\bibitem[Roslyak and Mukamel(2010)Roslyak, and Mukamel]{paper_663}
Roslyak,~O.; Mukamel,~S. Lectures of Virtual University, Max Born Institute,
  EVU Lecture Notes\relax
\mciteBstWouldAddEndPuncttrue
\mciteSetBstMidEndSepPunct{\mcitedefaultmidpunct}
{\mcitedefaultendpunct}{\mcitedefaultseppunct}\relax
\EndOfBibitem
\bibitem[Baym and Kadanoff(1961)Baym, and Kadanoff]{BaymKadanoffPR61}
Baym,~G.; Kadanoff,~L.~P. Conservation Laws and Correlation Functions.
  \emph{Phys. Rev.} \textbf{1961}, \emph{124}, 287--299\relax
\mciteBstWouldAddEndPuncttrue
\mciteSetBstMidEndSepPunct{\mcitedefaultmidpunct}
{\mcitedefaultendpunct}{\mcitedefaultseppunct}\relax
\EndOfBibitem
\bibitem[Baym(1962)]{BaymPR62}
Baym,~G. Self-Consistent Approximations in Many-Body Systems. \emph{Phys. Rev.}
  \textbf{1962}, \emph{127}, 1391--1401\relax
\mciteBstWouldAddEndPuncttrue
\mciteSetBstMidEndSepPunct{\mcitedefaultmidpunct}
{\mcitedefaultendpunct}{\mcitedefaultseppunct}\relax
\EndOfBibitem
\bibitem[Kadanoff and Baym(1962)Kadanoff, and Baym]{kadanoff_quantum_1962}
Kadanoff,~L.~P.; Baym,~G. In \emph{Quantum {Statistical} {Mechanics}};
  Pines,~D., Ed.; Frontiers in {Physics}; W. A. Benjamin, Inc.: New York,
  1962\relax
\mciteBstWouldAddEndPuncttrue
\mciteSetBstMidEndSepPunct{\mcitedefaultmidpunct}
{\mcitedefaultendpunct}{\mcitedefaultseppunct}\relax
\EndOfBibitem
\bibitem[Gao and Galperin(2016)Gao, and Galperin]{gao_optical_2016}
Gao,~Y.; Galperin,~M. Optical Spectroscopy of Molecular Junctions:
  {Nonequilibrium} {Green}'s Functions Perspective. \emph{J. Chem. Phys.}
  \textbf{2016}, \emph{144}, 174113\relax
\mciteBstWouldAddEndPuncttrue
\mciteSetBstMidEndSepPunct{\mcitedefaultmidpunct}
{\mcitedefaultendpunct}{\mcitedefaultseppunct}\relax
\EndOfBibitem
\bibitem[Nitzan and Galperin(2018)Nitzan, and Galperin]{nitzan_kinetic_2018}
Nitzan,~A.; Galperin,~M. Kinetic {Schemes} in {Open} {Interacting} {Systems}.
  \emph{J. Phys. Chem. Lett.} \textbf{2018}, \emph{9}, 4886--4892\relax
\mciteBstWouldAddEndPuncttrue
\mciteSetBstMidEndSepPunct{\mcitedefaultmidpunct}
{\mcitedefaultendpunct}{\mcitedefaultseppunct}\relax
\EndOfBibitem
\bibitem[Haug and Jauho(2008)Haug, and Jauho]{HaugJauho_2008}
Haug,~H.; Jauho,~A.-P. \emph{Quantum {K}inetics in {T}ransport and {O}ptics of
  {S}emiconductors}; Springer: Berlin Heidelberg, 2008\relax
\mciteBstWouldAddEndPuncttrue
\mciteSetBstMidEndSepPunct{\mcitedefaultmidpunct}
{\mcitedefaultendpunct}{\mcitedefaultseppunct}\relax
\EndOfBibitem
\bibitem[Stefanucci and van Leeuwen(2013)Stefanucci, and van
  Leeuwen]{StefanucciVanLeeuwen_2013}
Stefanucci,~G.; van Leeuwen,~R. \emph{Nonequilibrium Many-Body Theory of
  Quantum Systems. A Modern Introduction.}; Cambridge University Press,
  2013\relax
\mciteBstWouldAddEndPuncttrue
\mciteSetBstMidEndSepPunct{\mcitedefaultmidpunct}
{\mcitedefaultendpunct}{\mcitedefaultseppunct}\relax
\EndOfBibitem
\bibitem[Luttinger and Ward(1960)Luttinger, and Ward]{LuttingerWardPR60}
Luttinger,~J.~M.; Ward,~J.~C. Ground-State Energy of a Many-Fermion System. II.
  \emph{Phys. Rev.} \textbf{1960}, \emph{118}, 1417--1427\relax
\mciteBstWouldAddEndPuncttrue
\mciteSetBstMidEndSepPunct{\mcitedefaultmidpunct}
{\mcitedefaultendpunct}{\mcitedefaultseppunct}\relax
\EndOfBibitem
\bibitem[Haussmann(1999)]{Haussmann_1999}
Haussmann,~R. \emph{Self-consistent Quantum-Field Theory and Bosonization for
  Strongly Correlated Electron Systems}; Springer-Verlag: Berlin Heidelberg,
  1999\relax
\mciteBstWouldAddEndPuncttrue
\mciteSetBstMidEndSepPunct{\mcitedefaultmidpunct}
{\mcitedefaultendpunct}{\mcitedefaultseppunct}\relax
\EndOfBibitem
\bibitem[Stan \latin{et~al.}(2009)Stan, Dahlen, and van
  Leeuwen]{stan_time_2009}
Stan,~A.; Dahlen,~N.~E.; van Leeuwen,~R. Time propagation of the
  {Kadanoff}–{Baym} equations for inhomogeneous systems. \emph{J. Chem.
  Phys.} \textbf{2009}, \emph{130}, 224101\relax
\mciteBstWouldAddEndPuncttrue
\mciteSetBstMidEndSepPunct{\mcitedefaultmidpunct}
{\mcitedefaultendpunct}{\mcitedefaultseppunct}\relax
\EndOfBibitem
\bibitem[Latini \latin{et~al.}(2014)Latini, Perfetto, Uimonen, van Leeuwen, and
  Stefanucci]{latini_charge_2014}
Latini,~S.; Perfetto,~E.; Uimonen,~A.-M.; van Leeuwen,~R.; Stefanucci,~G.
  Charge dynamics in molecular junctions: {Nonequilibrium} {Green}'s function
  approach made fast. \emph{Phys. Rev. B} \textbf{2014}, \emph{89},
  075306\relax
\mciteBstWouldAddEndPuncttrue
\mciteSetBstMidEndSepPunct{\mcitedefaultmidpunct}
{\mcitedefaultendpunct}{\mcitedefaultseppunct}\relax
\EndOfBibitem
\bibitem[Jauho \latin{et~al.}(1994)Jauho, Wingreen, and
  Meir]{jauho_time-dependent_1994}
Jauho,~A.-P.; Wingreen,~N.~S.; Meir,~Y. Time-Dependent Transport in Interacting
  and Noninteracting Resonant-Tunneling Systems. \emph{Physical Review B}
  \textbf{1994}, \emph{50}, 5528--5544\relax
\mciteBstWouldAddEndPuncttrue
\mciteSetBstMidEndSepPunct{\mcitedefaultmidpunct}
{\mcitedefaultendpunct}{\mcitedefaultseppunct}\relax
\EndOfBibitem
\bibitem[Galperin \latin{et~al.}(2007)Galperin, Nitzan, and
  Ratner]{MGNitzanRatner_heat_PRB07}
Galperin,~M.; Nitzan,~A.; Ratner,~M.~A. Heat Conduction in Molecular Transport
  Junctions. \emph{Phys. Rev. B} \textbf{2007}, \emph{75}, 155312\relax
\mciteBstWouldAddEndPuncttrue
\mciteSetBstMidEndSepPunct{\mcitedefaultmidpunct}
{\mcitedefaultendpunct}{\mcitedefaultseppunct}\relax
\EndOfBibitem
\bibitem[Bergmann and Galperin(2019)Bergmann, and
  Galperin]{bergmann_electron_2019}
Bergmann,~N.; Galperin,~M. Electron {Transfer} {Methods} in {Open} {Systems}.
  \emph{J. Phys. Chem. B} \textbf{2019}, \emph{123}, 7225--7232\relax
\mciteBstWouldAddEndPuncttrue
\mciteSetBstMidEndSepPunct{\mcitedefaultmidpunct}
{\mcitedefaultendpunct}{\mcitedefaultseppunct}\relax
\EndOfBibitem
\bibitem[Leijnse and Wegewijs(2008)Leijnse, and Wegewijs]{leijnse_kinetic_2008}
Leijnse,~M.; Wegewijs,~M.~R. Kinetic Equations for Transport Through
  Single-Molecule Transistors. \emph{Phys. Rev. B} \textbf{2008}, \emph{78},
  235424\relax
\mciteBstWouldAddEndPuncttrue
\mciteSetBstMidEndSepPunct{\mcitedefaultmidpunct}
{\mcitedefaultendpunct}{\mcitedefaultseppunct}\relax
\EndOfBibitem
\bibitem[Koller \latin{et~al.}(2010)Koller, Grifoni, Leijnse, and
  Wegewijs]{koller_density-operator_2010}
Koller,~S.; Grifoni,~M.; Leijnse,~M.; Wegewijs,~M.~R. Density-Operator
  Approaches to Transport Through Interacting Quantum Dots: {Simplifications}
  in Fourth-Order Perturbation Theory. \emph{Phys. Rev. B} \textbf{2010},
  \emph{82}, 235307\relax
\mciteBstWouldAddEndPuncttrue
\mciteSetBstMidEndSepPunct{\mcitedefaultmidpunct}
{\mcitedefaultendpunct}{\mcitedefaultseppunct}\relax
\EndOfBibitem
\end{mcitethebibliography}

\providecommand{\latin}[1]{#1}
\makeatletter
\providecommand{\doi}
  {\begingroup\let\do\@makeother\dospecials
  \catcode`\{=1 \catcode`\}=2 \doi@aux}
\providecommand{\doi@aux}[1]{\endgroup\texttt{#1}}
\makeatother
\providecommand*\mcitethebibliography{\thebibliography}
\csname @ifundefined\endcsname{endmcitethebibliography}
  {\let\endmcitethebibliography\endthebibliography}{}

\begin{tocentry}
{\centering\includegraphics[width=\linewidth]{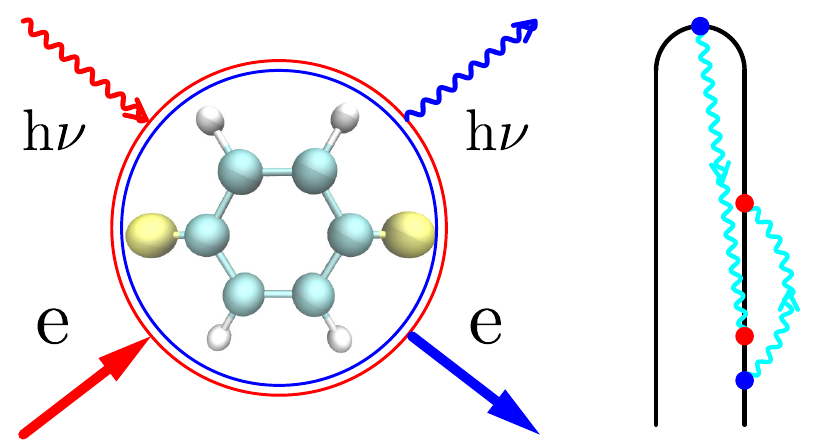}}\\
Conserving diagrammatic formulation of optical spectroscopy of open quantum systems.
\end{tocentry}

\end{document}